\documentclass[twocolumn,twocolappendix]{aastex63}

\usepackage[figuresright]{rotating}
\usepackage{amsmath}
\usepackage{amsfonts}
\usepackage{graphicx}
\usepackage{rotating}
\usepackage{natbib}
\usepackage{txfonts}
\usepackage{wrapfig}
\usepackage{amssymb}
\usepackage{color}
\usepackage{xcolor}
\usepackage{ulem}

\usepackage{lineno}
\linenumbers

\shorttitle{Aldehydes and their related alcohols}
\shortauthors{Mondal et al.}

\begin{document}
\title{Is there any linkage between interstellar aldehyde and
alcohol?}
\correspondingauthor{Ankan Das}
\email{ankan.das@gmail.com}
\author[0000-0002-7657-1243]{Suman Kumar Mondal}
\affiliation{Indian Centre for Space Physics, 43 Chalantika, Garia Station Road, Kolkata 700084, India}
\author[0000-0003-1602-6849]{Prasanta Gorai}
\affiliation{Department of Space, Earth \& Environment, Chalmers University of Technology, SE-412 96 Gothenburg, Sweden}
\affiliation{Indian Centre for Space Physics, 43 Chalantika, Garia Station Road, Kolkata 700084, India}
\author[0000-0001-5720-6294]{Milan Sil}
\author[0000-0003-1745-9718]{Rana Ghosh}
\affiliation{Indian Centre for Space Physics, 43 Chalantika, Garia Station Road, Kolkata 700084, India}
\author{Emmanuel E. Etim}
\affiliation{Department of Chemical Sciences, Federal University Wukari, Katsina-Ala Road, P.M.B. 1020 Wukari, Taraba State, Nigeria}
\author[0000-0002-0193-1136]{Sandip K Chakrabarti}
\affiliation{Indian Centre for Space Physics, 43 Chalantika, Garia Station Road, Kolkata 700084, India}
 \author{Takashi Shimonishi}
\affiliation{Center for Transdisciplinary Research,
Niigata University, Nishi-ku, Niigata 950-2181, Japan}
\affiliation{Environmental Science Program, Department of Science, Faculty
of Science, Niigata University, Nishi-ku, Niigata 950-2181, Japan}
\author{Naoki Nakatani}
\affiliation{Institute for Catalysis, Hokkaido University, N21W10 Kita-ku, Sapporo, Hokkaido 001-0021, Japan \&
Department of Chemistry, Graduate School of Science and Engineering, Tokyo Metropolitan University, 1-1 Minami-Osawa, Hachioji Tokyo 192-0397, Japan}
\author{Kenji Furuya}
\affiliation{Center for Computational Sciences, University of Tsukuba, Tsukuba, 305-8577, Japan \& National Astronomical Observatory of Japan, Tokyo 181-8588, Japan}
\author[0000-0002-3389-9142]{Jonathan C. Tan}
\affiliation{Department of Space, Earth \& Environment, Chalmers University of Technology, SE-412 96 Gothenburg, Sweden}
\affiliation{Department of Astronomy, University of Virginia, Charlottesville, VA 22904, USA}
\author[0000-0003-4615-602X]{Ankan Das}
\affiliation{Indian Centre for Space Physics, 43 Chalantika, Garia Station Road, Kolkata 700084, India}

\begin{abstract}
It is speculated that there might be some linkage between interstellar aldehydes and their corresponding alcohols. Here, an observational study and astrochemical modeling  are coupled together to illustrate the connection between them. The ALMA cycle 4 data of a hot molecular core, G10.47+0.03 is utilized for this study. Various aldehydes (acetaldehyde, propanal, and glycolaldehyde), alcohols (methanol and ethylene glycol), and a ketone  (acetone) are identified in this source. The excitation temperatures and the column densities of these species were derived via the rotation diagram method assuming LTE conditions.
An extensive investigation is carried out to understand the formation of these species.
Six pairs of aldehyde-alcohol:
i) methanal and methanol; ii) ethanal and ethanol;
iii) propanal and 1-propanol; iv) propenal and allyl alcohol; v) propynal and propargyl alcohol; vi) glycolaldehyde and ethylene glycol;
vii)  along with one pair of ketone-alcohol
(acetone and isopropanol) and viii) ketene-alcohol (ethenone and vinyl alcohol) are considered for this study.
Two successive hydrogenation reactions in the ice phase are examined to form these alcohols
from aldehydes, ketone, and ketene, respectively.
Quantum chemical methods are extensively executed to review the ice phase formation route and the kinetics
of these species. Based on the obtained kinetic data, astrochemical modeling is employed to derive the abundances of these aldehydes, alcohols, ketone, and ketene in this source. It is seen that our model could successfully explain the observed abundances of various species in this hot molecular core.
\end{abstract}


\section{Introduction} \label{sec:intro}

More than 200 molecular species have been identified in the interstellar medium (ISM) and circumstellar shells  \citep[][\url{https://cdms.astro.uni-koeln.de/classic/molecules}]{mcgu18}, which resolved
the much-unexplained mystery of the molecular Universe. 
Interstellar grains accelerate the formation of
Complex Organic Molecules (COMs, molecule with $>$ 6 atoms) in space. During the warm-up phase of a star-forming core, radicals or simple molecules on the grain surface become mobile and could produce various COMs \citep{chak06a,chak06b,garr06, garr08, das08a,das10, das11, das16, sil18}.
These ice phase species may transfer to the gas phase by various desorption mechanisms.

Alcohols and aldehydes were identified in various parts of the ISM. 
 Methanal (formaldehyde, $\rm{H_2CO}$) is the simplest form of aldehydes, which was first observed in space by \cite{snyd69}. It is an intermediary of the large complex organic molecules, which can help constrain physical conditions in the star-forming regions \citep{pers18}. Methanol (methyl alcohol, $\rm{CH_3OH}$) is the simplest alcohol and is one of the most abundant interstellar molecules. It was widely observed toward various sources \citep{allam92,pont03} and is used as a reliable tracer of high-density environments \citep{ment88}. A wide number of studies reported the chemical origin of methanol starting from formaldehyde \citep{wata02,fuch09,goum11,song17}. Methanol is ubiquitous in star-forming regions \citep{weav18} and is mainly formed on the interstellar dust via successive hydrogen addition reactions with carbon monoxide \citep{fedo15a,chua16,buts17}.

\cite{woon02a} carried out a theoretical study of the formation of H$_2$CO and CH$_3$OH. Quantum chemical calculations were carried out to determine the activation barrier for the hydrogen addition reactions ($\rm{H+CO}$ and $\rm{H+H_2CO}$) in gas, with water clusters $\rm(H_2O)_n$ (n$\le3$) and  in
ice phases. He included explicit water molecules to show the catalytic effect on the activation barrier by surrounding water molecules. \cite{rimo14,das08b} studied methanol formation by successive hydrogen addition with CO in water ice. They compared their simulated abundance with  observations. Recently, \cite{song17} derived the tunneling rate constant of H$_2$CO + H reaction on amorphous solid water for various product channels: CH$_2$OH,
CH$_3$O, and H$_2$ + HCO. They also provided the activation barrier of these three product channels and rate coefficients.

 Identification of ethanal (CH$_3$CHO, acetaldehyde) was first reported by \cite{gott73} toward Sgr B2 and Sgr A. They reported an emission at $1065$ MHz, which was assigned to the $1_{10} \rightarrow 1_{11}$ transition of CH$_3$CHO. This assignment was  based on the agreement in radial velocity with other molecules, which were observed in the galactic center region.
 Then \cite{four74b} observed it toward Sgr B2 with the Parkes 64m telescope. Its successor, ethanol ($\rm{C_2H_5OH}$, ethyl alcohol), is among the earliest complex molecules, which was identified toward the Sgr B2 \citep{zuck75}. 

Propanal ($\rm{CH_3CH_2CHO}$, propionaldehyde) is identified in Sgr B2(N) \citep{holl04}, where it coexists with propynal (HC$_2$CHO) and propenal (CH$_2$CHCHO).
The detection of these three aldehydes toward the same source suggests that successive hydrogen additions might be an
efficient process to establish the chemical linkage between propynal, propenal,
and propanal \citep{holl04}.
Propenal (also known as acrolein)  is the simplest unsaturated aldehyde. It is the second most stable isomer of the $\rm{C_3H_4O}$ isomeric group \citep{etim18}.
Allyl alcohol (CH$_2$CHCH$_2$OH) is an isomer of propanal and the corresponding alcohol of propenal \citep{zave18}.
Acetaldehyde is supposed to be a potential ancestor of carbohydrates and propenal (also known as acrolein) \citep{pizz04,cord05}.

The corresponding alcohol of propanal is 1-propanol
($\rm{CH_3CH_2CH_2OH}$), which is yet to be observed in space. However, trans-ethyl methyl ether ($\rm{CH_3CH_2OCH_3}$), which is one of the isomers of the same isomeric group, was tentatively detected toward Orion KL \citep{terc15}.
\cite{zave18} reported that the propanal to 1-propanol production by the hydrogenation
reaction might be inefficient due to the efficient hydrogen abstraction reaction.

 The simplest form of sugar, glycolaldehyde (HOCH$_2$CHO), plays an essential role in forming amino acids and complex sugars \citep{webe98,coll95}. Glycolaldehyde was also detected outside of the Galactic Center towards the hot core G31.41+0.31 \citep{belt09} (hereafter, G31). Ethylene glycol (HOCH$_2$CH$_2$OH) is the corresponding alcohol of glycolaldehyde. It was detected in a wide variety of sources such as in the Galactic Center \citep{holl02, bell13},  NGC 1333-IRAS 2A \citep{maur14}, Orion bar \citep{brou15} and also in the comets C/1995 O1 (Hale-Bopp) \citep{crov04}, C/2012 F6 (Lemmon) and  C/2013 R1 (Lovejoy) \citep{bive14}. It was also identified in the Murchinson and Murray meteorites \citep{coop01}. The abundance ratio of glycolaldehyde and ethylene glycol in star-forming regions and comets can help understand the possible preservation of  complex organic molecules \citep{cout18}. Methyl formate (CH$_3$OCHO) and acetic acid (CH$_3$COOH) are two isomers of glycolaldehyde. The linkage between the methyl formate, acetic acid, and glycolaldehyde can constrain the understanding of the chemical and physical conditions of the star-forming core \citep{wood13}.
Acetone ($\rm{CH_3COCH_3}$) is one of the most important
molecules in organic chemistry, and it was the first species containing $10$ atoms identified
in the ISM \citep{comb87}. 
 Acetone was also found to be present in the comet 67P/Churyumov-Gerasimenko \citep{altw17}. 
Ethenone (CH$_2$CO, ketene) was identified for the first time toward Sgr B2(OH) by \cite{turn77} through microwave measurements.  Recently, \cite{turn20} perform experiment combined with high-level quantum
chemical calculations and provide compelling evidence of ketene formation in the processed  mixture of water and carbon monoxide ices explaining the observed ketene detection in deep space.
The simplest enol compound, vinyl alcohol (CH$_2$CHOH, ethenol), was identified in an interstellar
cloud of dust and gas near the center of the Milky Way galaxy toward Sgr B2(N) utilizing its millimeter-wave rotational transitions \citep{turn01}.
In the dark cloud, the non-energetic process plays an active role in processing the ice constituents. In contrast, in the latter stages of star and planet formation, various energetic procedures play a crucial role \citep{fedo17}.
Laboratory  experiments
of methanol dissociation studied through various energetic processes, such as photons, electrons, Vacuum Ultraviolet (VUV) photons, X-ray, etc. that could lead to the
synthesis of various COMs. \cite{ober09} found that the ratio between ethylene glycol to ethanol and methyl formate to ethanol varies with the ice temperature and
irradiation.

In this work, ALMA archival data of a hot molecular core, G10.47+0.03 is analyzed which is located at a distance of 8.6 kpc \citep{sann14}. 
This is a young cluster \citep{cesa98, rolf11} whose luminosity is $\sim \rm{5\times 10^{5}\ L_{\odot}}$ \cite{cesa10}. Many simple species, complex organics, and molecules with higher upper energy state (such as HC$_3$N, HCN) were observed in this source, suggesting an active site for the chemical enrichment \citep{iked01, wyro99, gibb04, rolf11, gora20}.

Recently \cite{qasi19} showed that the aldehyde and primary alcohol may be linked via successive hydrogen addition reactions. A similar connection is also observed between ketones and secondary alcohols, i.e., isopropanol ($\rm{CH_3CHOHCH_3}$)
can be synthesized via two successive hydrogen additions with acetone \citep{hiro98}. 
Here, several aldehydes and alcohols are identified. Obtained column density, rotation temperature, spatial distributions of these molecules are discussed to understand the morphological correlation between these aldehydes and alcohols. The astrochemical model is implemented to compare with the observational finding. This work demonstrates that consecutive hydrogen addition could contribute to forming some complex interstellar molecules. Here, the kinetics of the consecutive hydrogen addition reactions of alcohol production from aldehyde, ketone, and ketene are examined. Binding energies (BEs) of a species with a grain substrate are essential for constructing a chemical model. Most COMs are primarily produced on the dust surface and further desorbed back to the gas phase \citep{requ06}. Most of the time, BE of these species is estimated, which may induce uncertainties in the results. Our estimated BE values with the water ice surface are included in our chemical model to see their impact. The proton affinity (PA) is not directly included in our model. But it is computed to check the reactivity of these species with H$^+$.
The role of PA of various species and their intermediate steps is also discussed to synthesize new COMs. Some of our obtained kinetic parameters are directly included in our chemical model to compare between the observation and modeling yields.

The remainder of this paper is organized as follows. In Section \ref{sec:obs}, the observational results are presented. The computational details adopted to explain the chemical linkage between aldehyde
and alcohol are given in Section \ref{sec:kinetics}. Astrochemical modeling and its implication are discussed in Section \ref{sec:chemical_model} and finally
in Section \ref{sec:conclusion}, we conclude.

\section{Observations of a hot molecular core, G10.47+0.03} \label{sec:obs}

\subsection{Observations and data analysis}
This paper uses the ALMA cycle 4 archival data (12m array data) of the observations ($\#$2016.1.00929.S) of G10.47+0.03 (hereafter, G10). 
The pointing center of the G10 observations is located at $\alpha$(J2000) = 18$^\mathrm{h}$08$^\mathrm{m}$38$\fs$232, $\delta$ (J2000) =-19$^\circ$51$\arcmin$50$\farcs$4.
This observation was performed with band 4  having four spectral bands covering the sky frequencies of 
$129.50 - 131.44$ GHz, $147.50 - 149.43$ GHz, $153.00 - 154.93$ GHz,  and $158.49 - 160.43$ GHz and corresponding  angular resolution at four different frequencies are $1.67 ^{''}$ (14362 AU), $1.52^{''}$ (13072 AU), $1.66^{''}$ (14276 AU), and $1.76^{''}$ (15136 AU) respectively. The spectral resolution of the data is 976 kHz. The number of antennas used for the corresponding four spectral windows are 39, 40, 41 and 39, respectively.
The flux calibrator, phase calibrator, and bandpass calibrator of this observation were J1733-1304, J1832-2039, and J1924-2914. The systemic velocity of the source is $68$ km s$^{-1}$ \citep{rolf11}. A detailed description of the
observations was presented in \cite{gora20}.
Here, all the analysis (spectral and line analysis) are carried out by using CASA 4.7.2 software \citep{mcmu07}. Since G10 is a line-rich source, the corrected sigma-clipping method of the STATCONT package \citep{sanc17} are used to determine the continuum level in data cube. Further, the spectral window of the data cube is divided into two data cubes (continuum and line emission), using the `uvcontsub' command in the CASA program.
 Line identification is performed with the CASSIS  \citep{vast15} (This  software has been developed by IRAP-UPS/CNRS (\url{http://cassis.irap.omp.eu}), which uses the Cologne Database for Molecular Spectroscopy  
  \citep[CDMS\footnote{\url{https://cdms.astro.uni-koeln.de}},][]{mull01,mull05,endr16} and  Jet  Propulsion  Laboratory \citep[JPL\footnote{\url{https://spec.jpl.nasa.gov}},][]{Pick98} databases. Various physical parameters ($\rm V_{LSR}$, line blending, upper-state energy, Einstein coefficient, and 3 sigma significance level) are considered. LTE synthetic spectra are generated with the best-fitted parameters to compare the observed spectral feature and line identification. We have also used the \url{https://splatalogue.online} for the additional confirmation of our choices.

\startlongtable
\begin{deluxetable*}{lccccccccc}
\tablecaption{Summary of the line parameters of the observed molecules toward G10.47+0.03. \label{tab:dataobs}}
\tablewidth{0pt}
\tabletypesize{\footnotesize}
\tablehead{
\colhead{Species} & \colhead{${\rm J^{'}_{K_a^{'}K_c^{'}}}$-${\rm J^{''}_{K_a^{''}K_c^{''}}}$} & \colhead{Frequency} & \colhead{E$_u$} & \colhead{$\Delta$V} & \colhead{I$_{max}$} & \colhead{V$_{LSR}$} & \colhead{S$\mu^{2}$} & \colhead{$\int{T_{mb}}dV$}\\
\colhead{ } & \colhead{ } & \colhead{(GHz)} & \colhead{(K)} & \colhead{(km s$^{-1}$)} &\colhead{(K)}&\colhead{(km s$^{-1}$)} & \colhead{(Debye$^{2}$)} & \colhead{(K km s$^{-1}$)}}
\startdata
Methanol &$\rm{21_{-0,21}-21_{1,21}}$E, vt=0&129.720384&546.2&9.6$\pm$0.2&30.1$\pm$0.6&66.9$\pm$0.1&19.8&308.7$\pm$14.5\\
($\rm{CH_3OH}$)&$\rm{17_{-4,13}-18_{-3,16}}$E, vt=0&131.134094&450.9&11.1$\pm$0.2&45.7$\pm$0.5&66.6$\pm$0.5&21.8&540.0$\pm$17.7\\
&$\rm{7_{-1,7}-6_{0,6}}$E, vt=1&147.943673&356.3&11.1$\pm$0.8&32.1$\pm$1.9&66.2$\pm$0.4&14.4&378.5$\pm$52.8\\
&$\rm{15_{-0,15}-15_{1,15}}$E, vt=0& 148.111993&290.7&10.5$\pm$0.6&44.7$\pm$1.5&66.5$\pm$0.2&29.02&500.0$\pm$47.2\\
&$\rm{8_{-7,1}-7_{-6,1}}$E, vt=1&153.128697&664.5&10.4$\pm$0.2&9.15$\pm$0.1&66.8$\pm$0.1&15.9&101.1$\pm$3.1\\
&$\rm{12_{-0,12}-15_{1,12}}$E, vt=0&153.281282$^*$&193.8&11.37$\pm$0.2&40.28$\pm$0.6&66.3$\pm$0.1&30.7&487.4$\pm$14.5\\
&$\rm{11_{-0,11}-11_{1,11}}$ E, vt=0&154.425832$^*$&166.1&10.6$\pm$0.3&36.02$\pm$0.7&66.64$\pm$0.1&30.5&408.0$\pm$18.7\\
&&&&&&&&\\
Ethylene Glycol
&$\rm{33_{10,23}-32_{11,22}}$, 1-1&130.115657&323&7.7$\pm$0.07&2.55$\pm$0.01&67.8$\pm$	0.03&53&20.9$\pm$0.3\\
(CH$_2$OH)$_2$&$\rm{13_{9,4}-12_{9,3}}$, 0-1&130.998583&83.7&7.9$\pm$0.5&5.6$\pm$0.24&68.2$\pm$0.178&98.4&48.1$\pm$5.02\\
&$\rm{28_{7,21}-28_{6,23}}$, 1-0&131.229156&223.4&7.8$\pm$0.1&4.8$\pm$0.02&68.8$\pm$	0.04&158.3&39.8$\pm$0.7\\
&$\rm{27_{3,25}-27_{2,26}}$, 0-0&148.082465&185.5&7.2$\pm$0.13&4.8$\pm$0.15&69.3$\pm$0.12&87.3&36.89$\pm$1.8\\
&$\rm{39_{8,32}-39_{7,33}}$, 1-1&153.325448&414&10.5$\pm$0.5&3.9$\pm$0.1&69.4$\pm$0.15&385.1&44.14$\pm$3.13\\
&$\rm{8_{4,5}-7_{3,5}}$, 1-0&153.567383&25.4&8.9$\pm$0.27&13.8$\pm$0.33&66.9$\pm$0.11&60.4&131.45$\pm$7.08\\
&$\rm{13_{3,11}-12_{2,10}}$, 0-0&159.768239&48.8&8.73$\pm$0.25&9.32$\pm$0.15&68.4$\pm$0.08&62.8&86.65$\pm$3.87\\
&&&&&&&&\\
Acetaldehyde&$\rm{8_{5,4}-7_{5,3}}$,A&154.161467&89.8&7.3$\pm$0.7&5.0$\pm$0.4&67.6$\pm$0.3&61.7&38.8$\pm$0.10\\
($\rm{CH_3CHO}$)&$\rm{8_{3,5}-7_{3,4}}$,A&154.173895&114.4&6.6$\pm$0.2&1.3$\pm$0.1&67.8$\pm$0.1&44.3&9.2$\pm$0.53\\
&$\rm{8_{4,4}-7_{4,3}}$,A&154.201471&69.5&8.8$\pm$0.5&5.7$\pm$0.2&67.0$\pm$0.2&75.9&52.9$\pm$0.08\\
&$\rm{8_{3,6}-7_{3,5}}$,A&154.274686&53.7&8.4$\pm$0.2&5.3$\pm$0.1&67.1$\pm$0.7&87.0&47.5$\pm$0.01\\
&$\rm{8_{3,5}-7_{3,4}}$,E&154.296489&53.7&8.3$\pm$0.2&4.5$\pm$0.1&66.8$\pm$0.1&87.0&39.6$\pm$0.03\\
&&&&&&&&\\
Propanal &$\rm{10_{7,4}-10_{6,5}}$, A&147.867697&54.6&12.1$\pm$1.2&12.3$\pm$0.8&68.9$\pm$0.4&12.0&158.4$\pm$6.5\\
($\rm{CH_3CH_2CHO}$)&$\rm{9_{7,3}-9_{6,4}}$, A&148.005214&49.6&10.1$\pm$0.3&11.5$\pm$0.3&67.6$\pm$0.1&9.3&123.8$\pm$1.5\\
&$\rm{25_{3,22}-25_{2,23}}$, A&153.554511&172.7&9.0$\pm$0.3	&13.2$\pm$0.3&68.9$\pm$0.1&31.2	&126.2$\pm$3.5\\
&$\rm{27_{5,23}-27_{4,24}}$, A&154.067527&206.2&8.5$\pm$0.3&13.5$\pm$0.3&69.8$\pm$0.1&41.6&123.4$\pm$3.3\\
&$\rm{16_{2,15}-15_{1,14}}$,A&159.932413&68.6&9.9$\pm$0.2&18.7$\pm$0.2&68.3$\pm$0.1&38.1&198.2$\pm$3.2\\
&&&&&&&&\\
Glycolaldehyde&$\rm{15_{0,15}-14_{1 14}}$&153.597996&60.5&8.1$\pm$0.3&3.9$\pm$0.1&66.9$\pm$0.1&72.9&11.7$\pm$2.3\\
($\rm{HOCH_2CHO}$)&$\rm{20_{7,13}-20_{6,14}}$&153.614231&147.0&7.1$\pm$0.6&2.1$\pm$0.2&66.4$\pm$0.2&60.4&11.5$\pm$2.5\\
&&&&&&&&\\
Acetone&$\rm{25_{10,15}-25_{9,16}}$, AE&130.708968&241&8.2$\pm$0.3&4.43$\pm$0.1&65.8$\pm$0.1&536.9&38.8$\pm$	2.1\\
($\rm{CH_3COCH_3}$)&$\rm{12_{2,11}-11_{1,10}}$, AA&130.924799&44.1&10.6$\pm$0.2&7.7$\pm$0.1&66.3$\pm$0.1&876.4&
86.6$\pm$2.9\\
&$\rm{11_{5,7}-10_{4,6}}$, EE&147.684364&48.4&9.6$\pm$0.2&6.13$\pm$0.1&67.5$\pm$0.1&783.8&62.7$\pm$2.5\\
&$\rm{26_{10,17}-26_{9,18}}$, EE&149.190766&251&10.9$\pm$0.7&7.4$\pm$0.3&65.5$\pm$0.2&1309.1&86.1$\pm$8.2\\
&$\rm{13_{2,11}-12_{3,10}}$, AE&149.395864&55.8&10.8$\pm$1.1&8.4$\pm$0.5&68.3$\pm$0.4&517.6&96.7$\pm$15.7\\
&$\rm{14_{2,12}-13_{3,11}}$, EE&159.247634&63&8.7$\pm$0.2&14.8$\pm$0.2&67.5$\pm$0.1&1516.9&137.3$\pm$5.5\\
&$\rm{24_{6,18}-24_{5,19}}$, EE&159.415955&198&8.8$\pm$0.2&9.6$\pm$0.1&68.4$\pm$	0.1&902.1&89.16$\pm$2.6\\
&$\rm{22_{4,18}-22_{3,19}}$, EE&160.118153&157.1&9.3$\pm$0.2&4.3$\pm$0.1&65.5$\pm$0.1&607.6&42.4$\pm$2.0\\
&&&&&&&&\\
Methyl formate &$\rm{11_{2,10}-10_{2,9}}$, A&130.016790&40.7&9.5$\pm$0.20&17.55$\pm$0.3&66.1$\pm$0.07&28.04&178.3$\pm$6.44\\
($\rm{CH_3OCHO}$)&$\rm{11_{2,10}-10_{2,9}}$, E&130.010105&40.7&9.4$\pm$0.17&17.14$\pm$0.12&66.4$\pm$0.04&28.04&171.7$\pm$4.46\\
&$\rm{12_{1,12}-11_{1,11}}$, A&131.377495&230&9.38$\pm$0.2&6.28$\pm$0.05&66.7$\pm$0.05&31.39&62.7$\pm$1.83\\
&$\rm{12_{4,8}-11_{4,7}}$, E&148.575217&244.1&8.65$\pm$0.9&8.13$\pm$0.73&66.9$\pm$0.4&28.3&74.87$\pm$14.48\\
&$\rm{12_{4,9}-11_{4,8}}$, A&148.805941&56.8&8.52$\pm$0.44&18.13$\pm$0.7&66.3$\pm$0.2&28.4&164.50$\pm$14.98\\
&$\rm{13_{3,11}-12_{3,10}}$, A&158.704392&59.6&9.4$\pm$0.2&24.29$\pm$0.4&66.2$\pm$0.1&32.57&243.05$\pm$9.17\\
&$\rm{13_{10,4}-12_{10,3}}$, A&159.662793&120.04&8.71$\pm$0.3&18.78$\pm$0.3&66.64$\pm$0.07&14.1&174.12$\pm$4.38\\
&$\rm{13_{7,6}-12_{7,5}}$, A&160.178942&86.2&9.03$\pm$0.25&16.05$\pm$0.22&66.4$\pm$0.08&24.5&154.27$\pm$6.30\\
&$\rm{13_{7,6}-12_{7,5}}$, A&160.193496&86.2&9.5$\pm$0.15&35.15$\pm$0.32&66.48$\pm$0.04&24.5&358.00$\pm$3.00\\
 &&&&&&&&\\
Dimethyl ether&$\rm{6_{1,6}-5_{0,5}}$, EE&131.405796&19.9&11.8$\pm$0.5&28.7$\pm$0.5&66.1$\pm$0.2&82.4&359.9$\pm$20.6\\
($\rm{CH_3OCH_3}$)&$\rm{9_{0,9}-8_{1,8}}$, EE&153.054818&40.4&9.4$\pm$0.2&26.5$\pm$0.3&66.8$\pm$0.1&126.4&264.02$\pm$6.5\\
&$\rm{24_{4,20}-24_{3,21}}$, EE&153.385902&297.5&12.4$\pm$0.4&11.5$\pm$0.1&67.2$\pm$0.1&379.5&151.6$\pm$3.6\\
&$\rm{11_{1,10}-10_{2,9}}$, EE&154.455083&62.9&10.6$\pm$0.2&11.6$\pm$0.1&66.1$\pm$0.1&68.8&130.3$\pm$3.7\\
\enddata
\tablecomments{$^*$ optically thick}
\end{deluxetable*}

\subsection{Synthesized images}
Continuum maps at $1.88$ mm, $1.94$ mm, $2.02$ mm, and $2.30$ mm were presented in \cite{gora20}. They obtained the corresponding synthesized beam and position angle 2.38$^{''} \times1.39^{''}$ and 77.84$^{\circ}$; 2.03$^{''}\times$1.47$^{''}$ and 
 73.50$^{\circ}$; 1.98$^{''}\times$1.57$^{''}$ and 64.28$^{\circ}$; and 2.44$^{''}\times$1.64$^{''}$ and 63.16$^{\circ}$, respectively. Others parameters of continuum images such as frequency, peak flux, integrated flux, and deconvolved beam size (FWHM) were presented in Table 2 of \cite{gora20}. The synthesized beam size
of this data cube was not sufficient to resolve the continuum
emission. Hydrogen column density and dust optical depth of each continuum map was presented in Table 4 of \cite{gora20}. The  average value of hydrogen column density is  $\sim$ 1.35$\times$10$^{25}$cm$^{-2}$ and dust optical depth is 0.136 implies that the source as optically thin \citep{gora20}.

\subsection{Line analysis}
The observed spectra obtained within the circular region having a diameter of $2.0^{''}$ ($\sim 17200$ AU) centered at RA (J2000) = (18$^\mathrm{h}$08$^\mathrm{m}$38$\fs$232), Dec (J2000) =  (-19$^\circ$51$\arcmin$50$\farcs$4) are continuum substracted for the analysis. The line width ($\Delta V$), the LSR velocity ($V_{LSR}$), and the integrated intensity ($\int T_{mb} dV$) of the each transition which is obtained after Gaussian fitting to the observed transition is noted. All the line parameters corresponding to a observed transition such as associated quantum numbers {${\rm J^{'}_{K_a^{'}K_c^{'}}}$-${\rm J^{''}_{K_a^{''}K_c^{''}}}$}, 
rest frequency ($\nu_0$), $\Delta V$, $\int T_{mb} dV$, upper state energy ($E_u$), $V_{LSR}$ are presented in Table \ref{tab:dataobs}.
Multiple transitions of methanol (CH$_3$OH), acetaldehyde (CH$_3$CHO), ethylene glycol [(CH$_2$OH)$_2$], glycolaldehyde (HOCH$_2$CHO), propanal (CH$_3$CH$_2$CHO), acetone (CH$_3$COCH$_3$), methyl formate (CH$_3$COH), and dimethyl ether (CH$_3$OCH$_3$) are identified. This is the first time, ethylene glycol and propanal are identified in G10. Only two lines of  glycolaldehyde are identified. Therefore, it is considered as the tentative detection in this source.
 The observed spectra of these species and fitted spectra are shown in the Appendix (see Figure \ref{fig:fit1}). The integrated intensity for the methanol, ethylene glycol, and glycolaldehyde is obtained by simple Gaussian fitting to each transition.
In the case of acetaldehyde, propanal, acetone, methyl formate, and dimethyl ether, torsional sub-states due to the internal rotation of the methyl group are noticed. For acetone and dimethyl ether, these sub-states are AA, EE, EA, and AE.
In the case of acetaldehyde, propanal, and methyl formate, A and E  sub-states are noticed. But these transitions are found to be blended with different
sub-states. These transitions were not resolved due to the low spectral
resolution of the present spectra. The integrated intensity is obtained with the Gaussian fitting and then dividing this integrated intensity according to their S$\mu^2$ values (obtained from \url{https://www.cv.nrao.edu/php/splat}). The transition having maximum intensity between the torsional sub-states was considered in the rotation diagram analysis to calculate the rotation temperature and column density.

\subsection{Spatial distribution of aldehydes and alcohols}
The integrated intensity distribution of CH$_3$OH, CH$_3$CHO, (CH$_2$OH)$_2$, HOCH$_2$CHO, CH$_3$CH$_2$CHO, CH$_3$COCH$_3$, CH$_3$COH, and CH$_3$OCH$_3$ are shown in Appendix (see Figures \ref{fig:maps1}--\ref{fig:maps4}). 
The source sizes are determined by using two-dimensional Gaussian fitting of the images.
The deconvolved beam size of the emitting region is estimated
as $\theta_s=\sqrt{\theta^2_{50}-\theta^2_{beam}}$, where $\theta_{50}=2\sqrt{A/\pi}$ is diameter of the circle whose area (A) 
enclosing $50\%$ line peak and $\theta_{beam}$ is the half-power width of the synthesized beam \citep{rivi17}. It is noticed that all the lines have their peak around the position of the continuum. The emitting diameter of the observed species is summarized in Table \ref{tab:dia}. It is noticed that the emitting diameters of different observed transitions are smaller than the beam size. So all these transitions are not well spatially resolved or, at best, marginally resolved.
 In the case of methanol, the emitting region varies from $1.23^{''}$ to $1.81 ^{''}$, and it is noticed that the emitting regions of methanol decrease with an increase of upper state energy. Therefore, methanol could trace the temperature distribution of the source. However, our data cannot provide much insight into the source's temperature distributions due to the low angular resolution. All transitions of acetaldehyde having upper state energy $<100$ K are detected from the $1.33^{''}-1.53 ^{''}$ region. The emitting region of dimethyl ether and methyl formate transitions having upper state energy $<100$ K varies between $1.83^{''}-1.95 ^{''}$  and $1.66^{''}- 1.95 ^{''}$, respectively. So, dimethyl ether and methyl formate transitions are found from roughly the similar region as acetaldehyde. But the transitions of propanal, glycolaldehyde, acetone, and ethylene glycol are detected from different positions.  It is observed that the transitions of propanal, glycolaldehyde, and ethylene glycol emissions are very compact toward the dust continuum except for some transitions.
These suggest the grain surface origin of these species. Emissions of dimethyl ether and methyl formate are found to be comparatively extended compared to the other species. 
The emitting region of ethylene glycol is small compared to the emission of all other COMs identified in this work. The emitting region of dimethyl ether is found to be more extensive compared to all other species.

\begin{deluxetable}{ccccc}
\tablecaption{Estimated rotation temperature and column density of the observed molecules in G10. In the last column, the obtained column densities are compared with other observations of G10. \label{tab:rottmp}}
\tablewidth{0pt}
\tabletypesize{\scriptsize} 
\tablehead{
\colhead{Molecule} & \colhead{T$_{rot}$[K]} & \colhead{N$_{tot}$ [cm$^{-2}$]$^a$} &
\colhead{ Literature N$_{tot}$ [cm$^{-2}$]}}
\startdata
CH$_3$OH&$206\pm66 $ &$(4.3\pm3.4)\times10^{18}$  &$9.0\times10^{18b}$ \\
(CH$_2$OH)$_2$ & $185\pm61 $& $(5.0\pm 2.1)\times10^{17} $ & \\
CH$_3$CHO & $72\pm11 $ &$(6.8\pm 1.1)\times10^{15}$ & $(2.1 \pm 0.7) \times10^{14c}$ \\
CH$_3$CH$_2$CHO & $234\pm42$ &$(2.4\pm0.23)\times10^{17}$ & \\
HOCH$_2$CHO & 155 & $1.3\times10^{16}$ & \\
CH$_3$COCH$_3$ & $224\pm41$&$(5.4\pm0.7)\times10^{17}$& $5.0\times10^{17b}$ \\
CH$_3$OCHO & $191\pm24$&$(9.1\pm 0.7)\times10^{17}$&7.0$\times10^{17b}$ \\
CH$_3$OCH$_3$ & $148\pm42$&$(1.5\pm0.4)\times10^{17}$& $1.5\times10^{18b}$ \\
\enddata
\tablecomments{ Estimated values have
$12-33$\% and $8-79$\% uncertainty for temperature and column density,
respectively.}
$^a$ This work,\\ 
$^b$ \cite[$\theta_b=0.28 - 4.27^{''}$, $\theta_s = 1.5^{''}$, $\rm{T_{ex} = 200}$ K]{rolf11},\\ 
$^c$ \cite[ $\theta_b = 16-26^{''},\ \theta_s = 20^{''}, \ \rm{T_{rot} = 30 \pm 4}$ K]{iked01}
\end{deluxetable}

\begin{figure*}[t]
\begin{minipage}{0.42\textwidth}
\includegraphics[width=\textwidth]{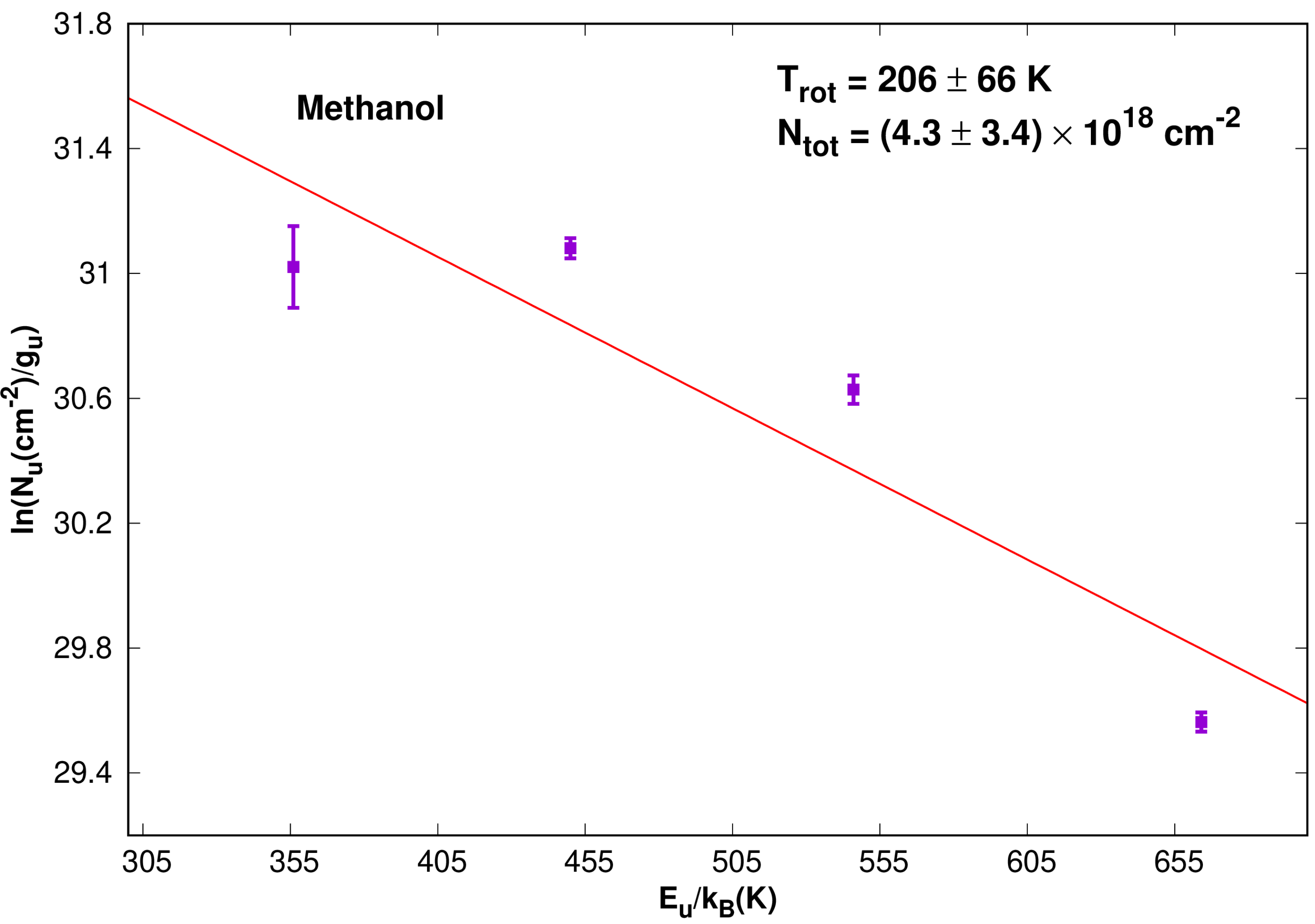}
\end{minipage}
\begin{minipage}{0.42\textwidth}
\includegraphics[width=\textwidth]{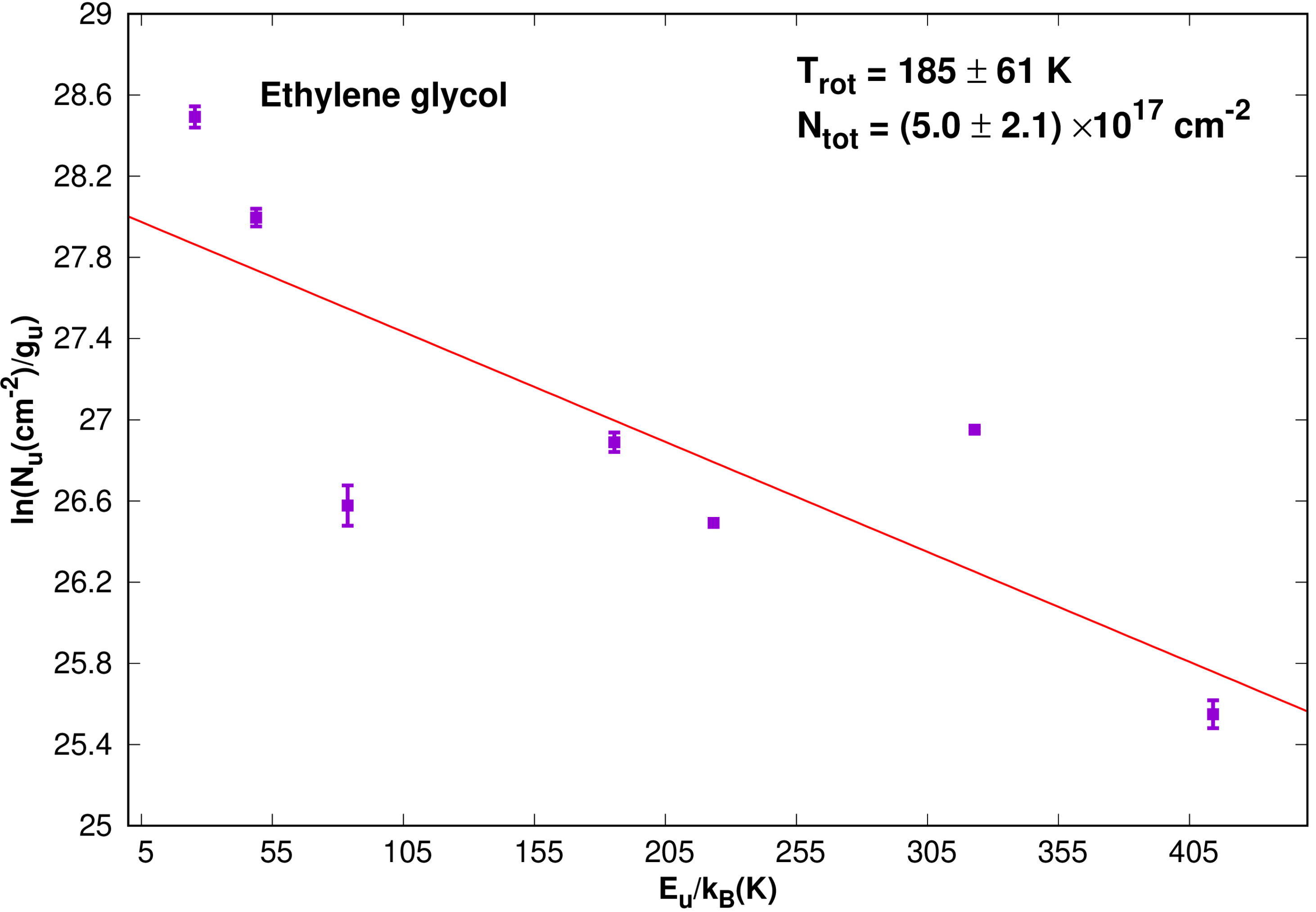}
\end{minipage}
\begin{minipage}{0.42\textwidth}
\includegraphics[width=\textwidth]{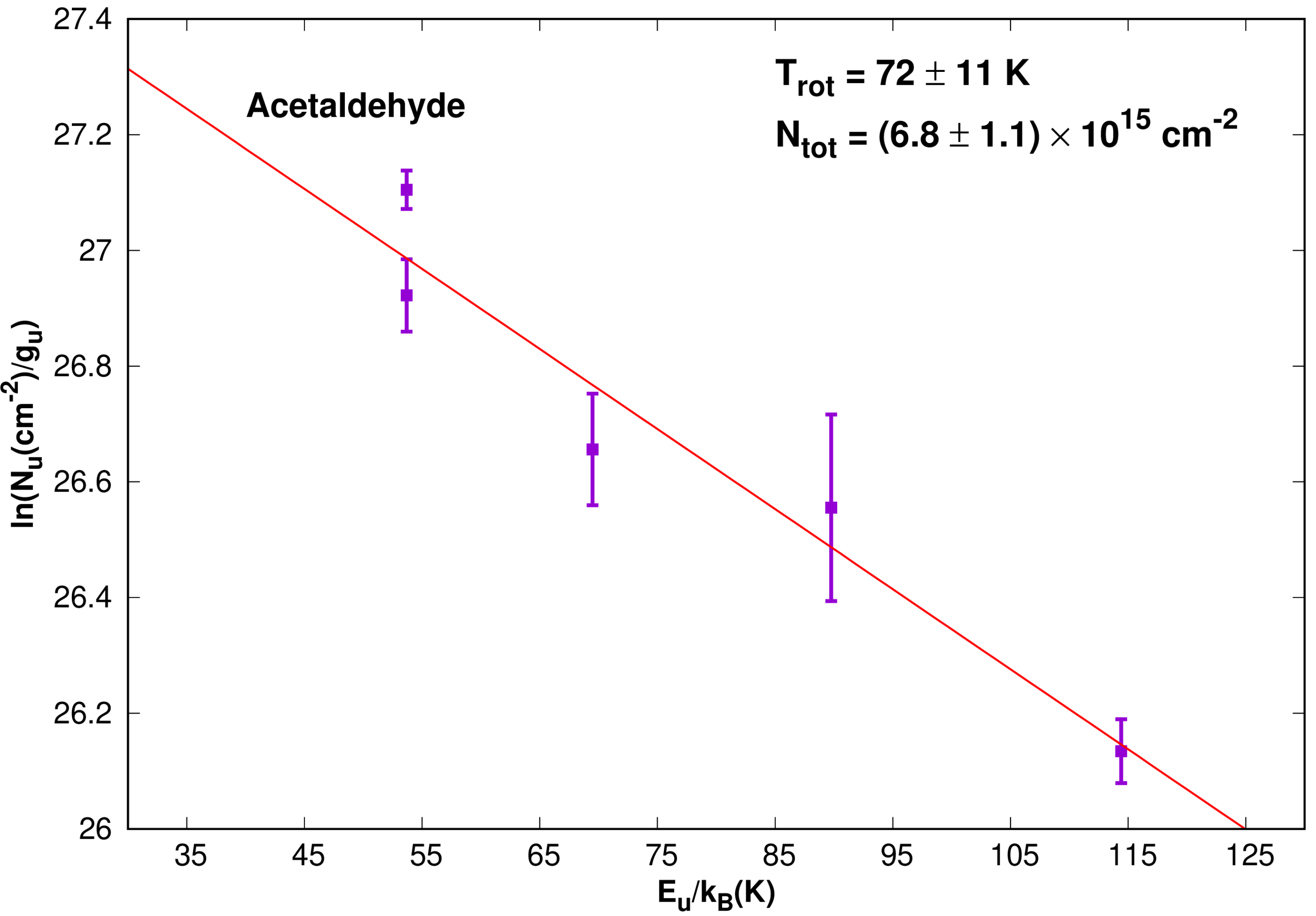}
\end{minipage}
\begin{minipage}{0.42\textwidth}
\includegraphics[width=\textwidth]{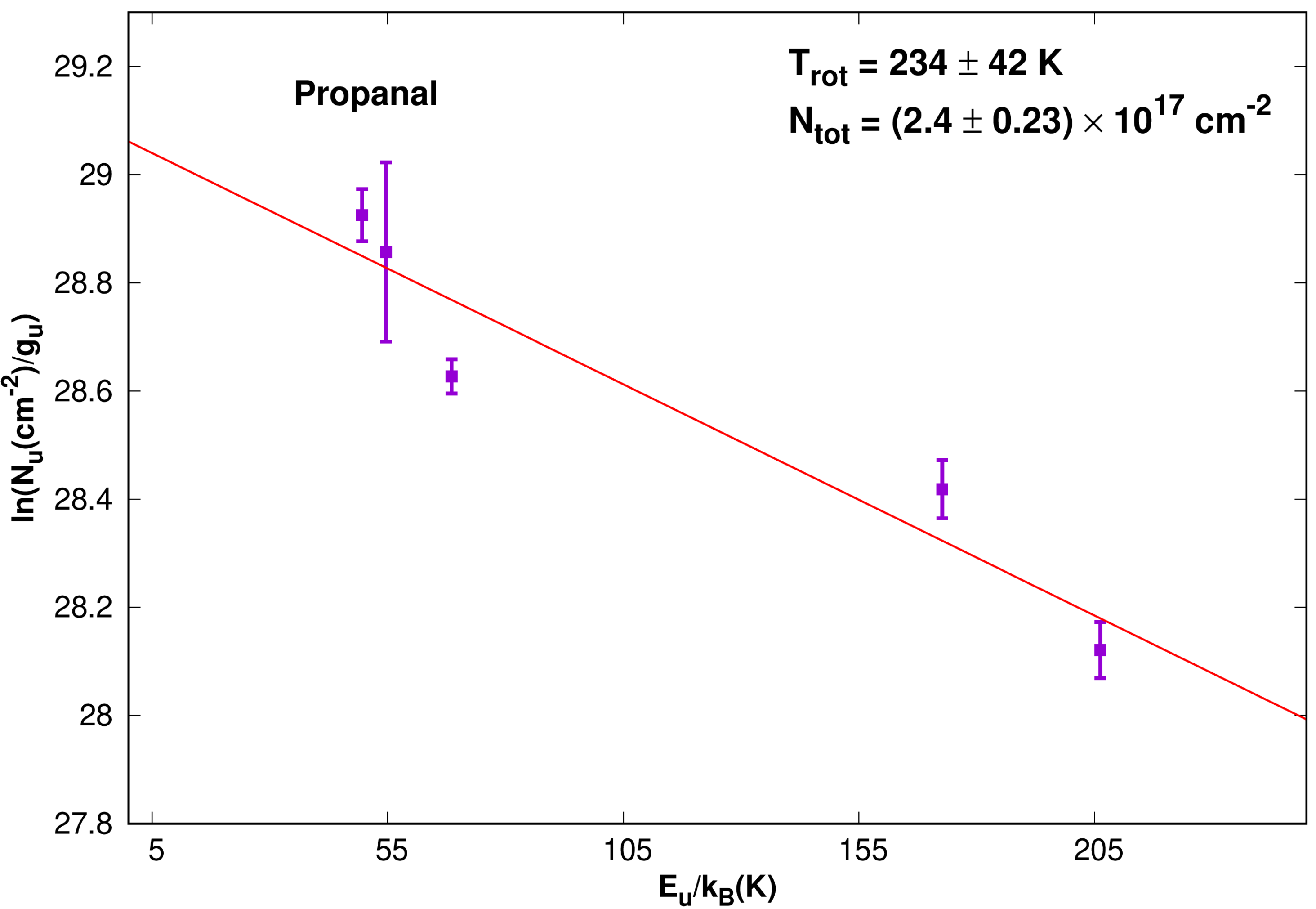}
\end{minipage}
\begin{minipage}{0.42\textwidth}
\includegraphics[width=\textwidth]{ga_rot.pdf}
\end{minipage}
\begin{minipage}{0.42\textwidth}
\includegraphics[width=\textwidth]{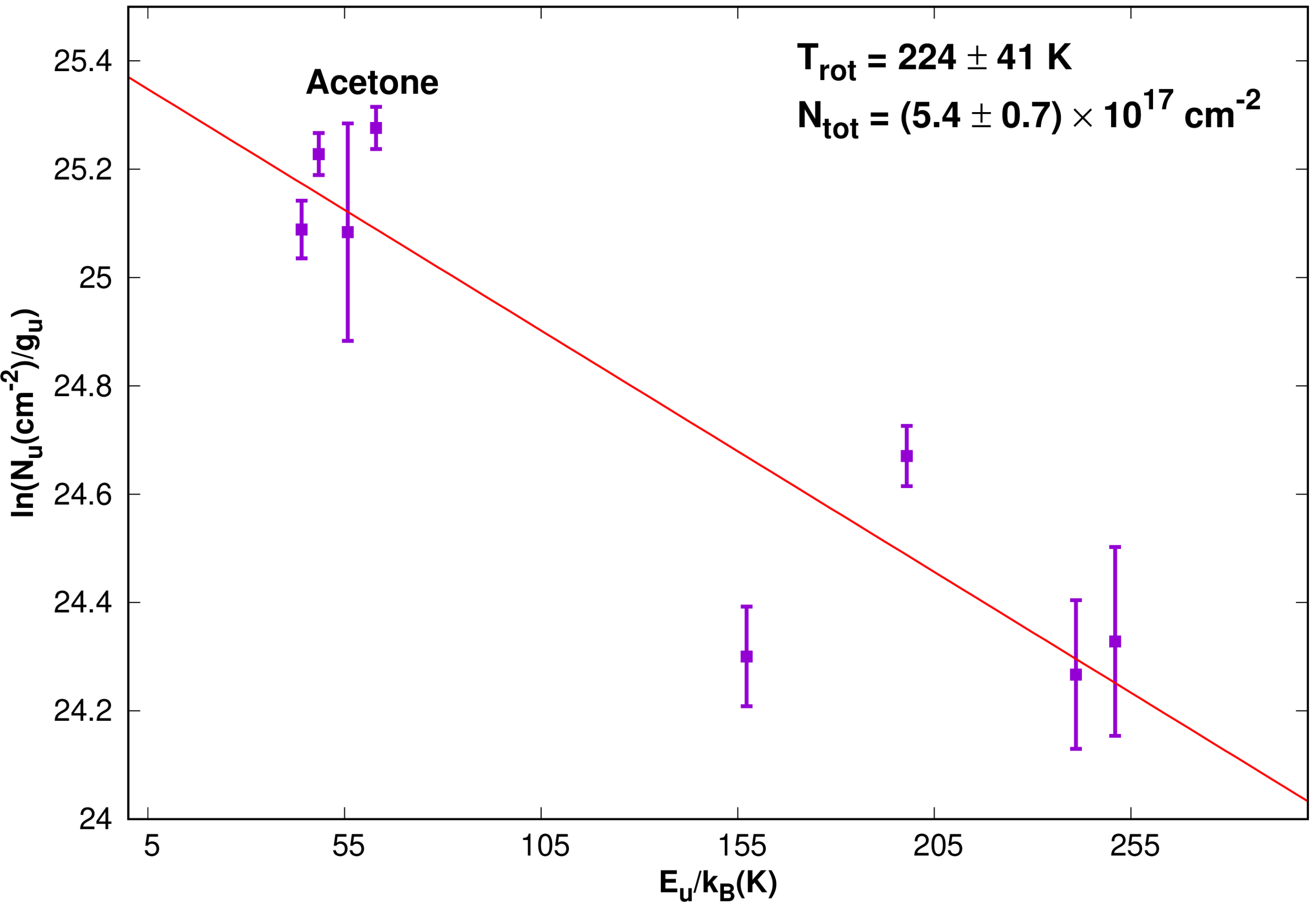}
\end{minipage}
\begin{minipage}{0.42\textwidth}
\includegraphics[width=\textwidth]{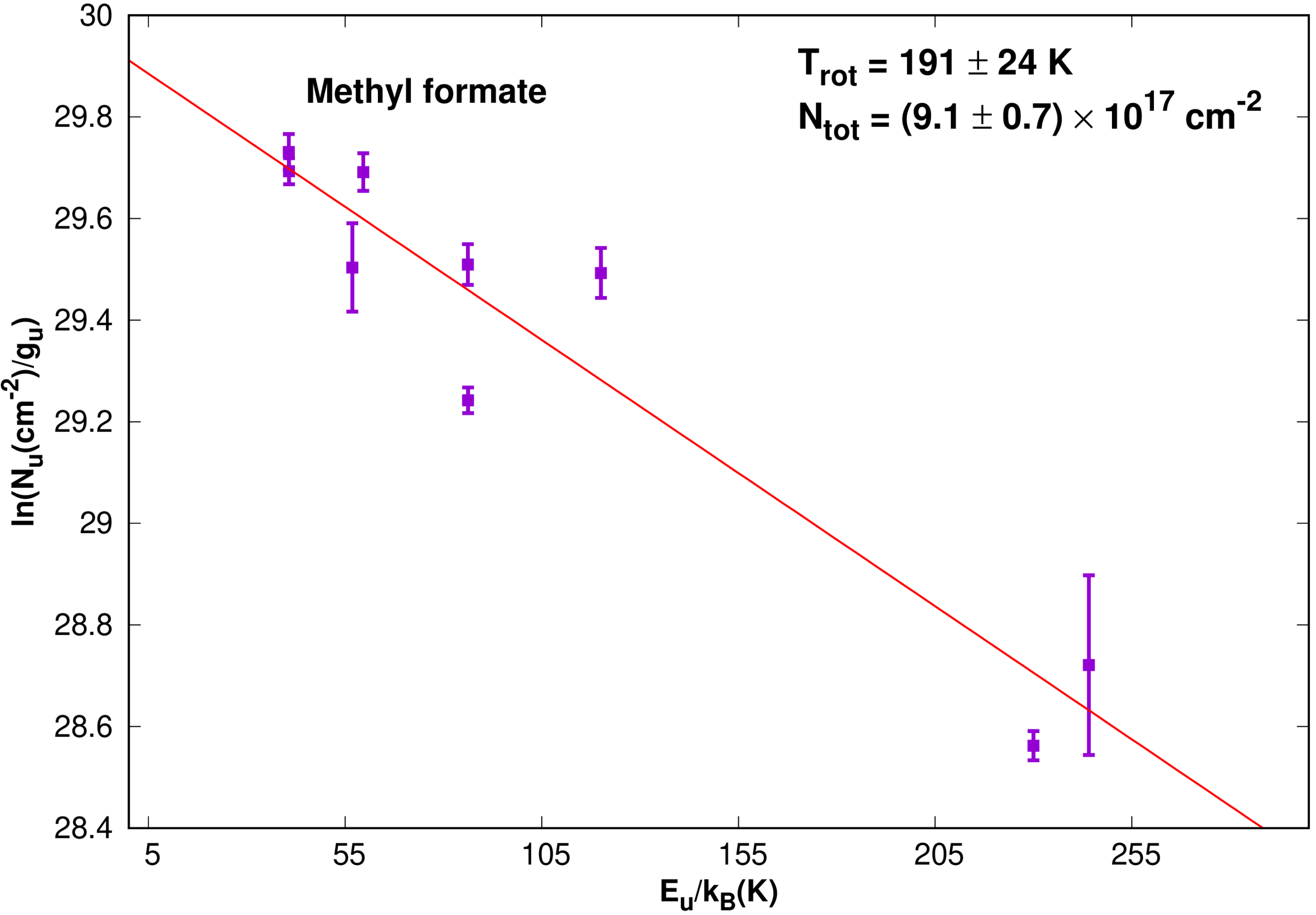}
\end{minipage}
 \hskip 2.8cm
\begin{minipage}{0.42\textwidth}
\includegraphics[width=\textwidth]{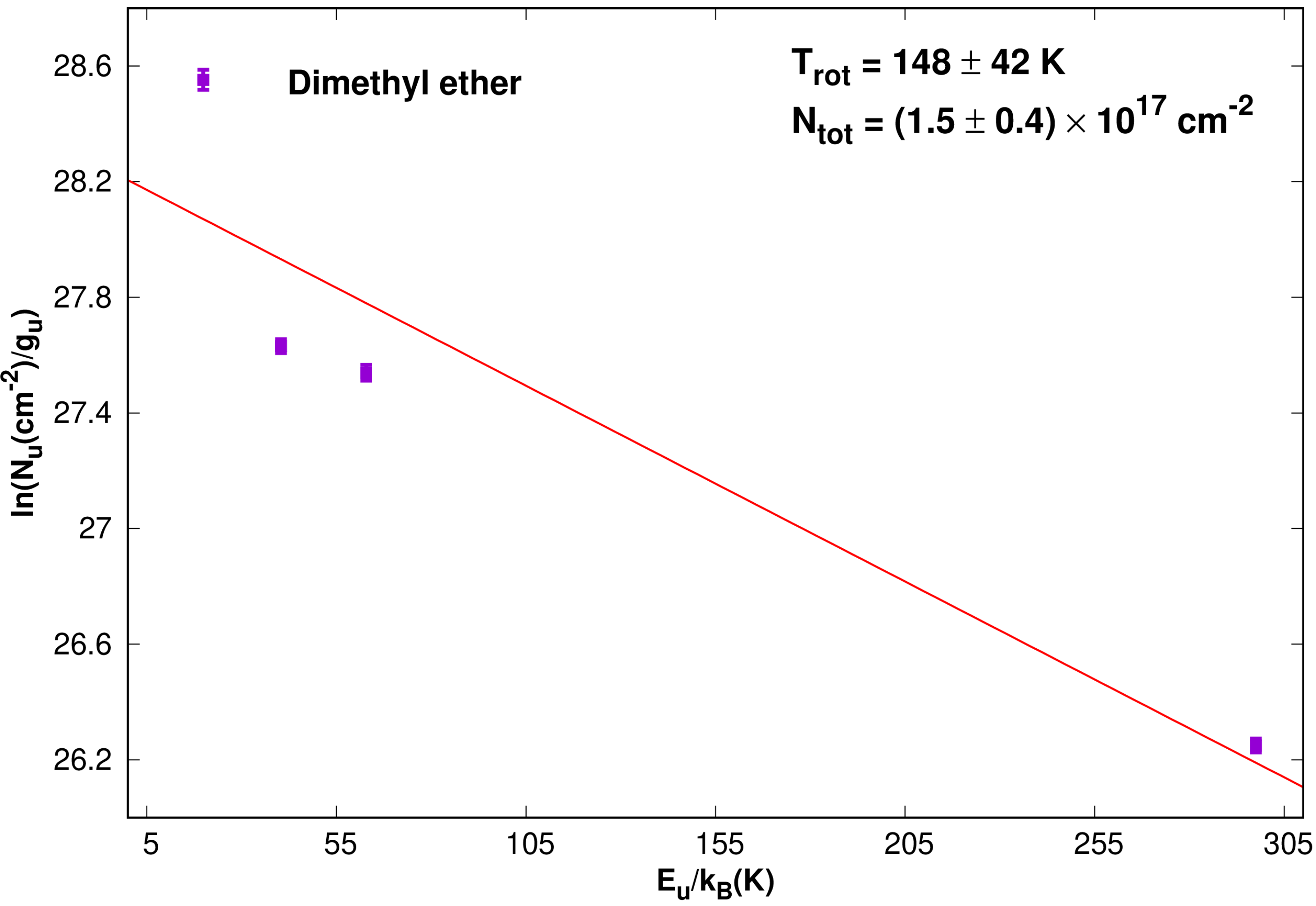}
\end{minipage}
\caption{Rotation diagram  of observed molecules are shown. The best fitted rotational temperature
and column density  are given in  each panel. Filled purple squares are the data points and purple line represents the error bar.}
\label{fig:rotdia}
\end{figure*}

\subsection{Rotation diagram analysis}
Here, a rotational diagram analysis is performed by assuming the observed transitions are optically thin and in
Local Thermodynamic Equilibrium (LTE). For optically thin lines, column density can be expressed as \citep{gold99},
\begin{equation}
\frac{N_u^{thin}}{g_u}=\frac{3k_B\int{T_{mb}dV}}{8\pi^{3}\nu S\mu^{2}},
\end{equation}
where, g$_u$ is the degeneracy of the upper state, $k_B$ is the Boltzmann constant, $\rm{\int T_{mb}dV}$ is the integrated intensity, $\nu$ is the
rest frequency, $\mu$ is the electric dipole moment, and S is the line strength. Under LTE condition, the total column density would be written as,
\begin{equation}
\frac{N_u^{thin}}{g_u}=\frac{N_{total}}{Q(T_{rot})}\exp^{-E_u/k_BT_{rot}},
\end{equation}
where, $T_{rot}$ is the rotational temperature, E$_u$ is the upper state energy, $\rm{Q(T_{rot})}$ is the partition function at rotational
temperature.  The above two equations can be
rearranged as,
\begin{equation}
ln\Bigg(\frac{N_u^{thin}}{g_u}\Bigg)=-\Bigg(\frac{1}{T_{rot}}\Bigg)\Bigg(\frac{E_u}{k_B}\Bigg)+ln\Bigg(\frac{N_{total}}{Q(T_{rot})}\Bigg).
\label{eqn:clmn}
\end{equation}

The rotational temperature and total column density of a species are simultaneously determined from this expression. All the spectroscopic parameters for this analysis are taken either from the 
Cologne Database for Molecular Spectroscopy \citep[CDMS, \url{https://cdms.astro.uni-koeln.de,}][]{mull01,mull05, endr16} or from the Jet Propulsion Laboratory datebase \citep[JPL, \url{http://spec.jpl.nasa.gov,}][]{Pick98}.
Rotational diagram of methanol (CH$_3$OH), acetaldehyde (CH$_3$CHO), ethylene glycol [(CH$_2$OH)$_2$], glycolaldehyde (HOCH$_2$CHO), propanal (CH$_3$CH$_2$CHO), acetone (CH$_3$COCH$_3$), methyl formate (CH$_3$OCHO), and dimethyl ether (CH$_3$OCH$_3$)
are shown in Figure \ref{fig:rotdia}. The derived rotational temperatures and column densities are summarized in Table \ref{tab:rottmp}. In Table  \ref{tab:rottmp}, we have noted the errors. The relative percentage errors of the rotational temperature and column density are noted at the footnote of the table.
The rotational diagram of glycolaldehyde is constructed with only two data points. Therefore, the rotational temperature has been assumed to some constant value. 
   Methanol transitions with a wide range (166 K - 664 K) of upper state energies are observed. Some of these transitions are optically thick (i.e., $153.281$ GHz and $154.425$ GHz). Therefore, it is challenging to construct the rotation diagram with all seven unblended methanol lines (noted in Table \ref{tab:dataobs}). Since two transitions are optically thick, these are excluded from the rotation diagram analysis. Furthermore, by considering the rest of the five transitions, considerable uncertainty in the estimated column density ($(3.7\pm  3) \times 10^{18}$ cm$^{-2}$) and rotational temperature ($345 \pm 198$ K) is obtained. Comparatively, a better linear fit was possible if we exclude the $148.111$ GHz transitions from the rotation diagram (by choosing the transitions having upper state energy $>300$ K). It yields a column density and rotational temperature of methanol ($4.3 \pm 3.4) \times 10^{18}$ cm$^{-2}$ and  $206 \pm 66$ K, respectively.

 For the sample of considered molecules, our estimated rotation temperature varies from $72$ K to $234$ K. A low rotation temperature $\sim 72 \pm 11$ K for acetaldehyde is obtained, which is in agreement with the rotation temperature of accetaldehyde obtained by \cite{iked01}. 
  \cite{olmi96} derived a temperature of 160 K for CH$_3$CN. \cite{rolf11} considered a temperature of 200 K to give an optimum fit to the data for line identification purposes. The average rotational temperature obtained from the species observed here is $177$ K, which is in close agreement with \cite{olmi96,rolf11}. 
  There could be a possibility of beam dilution on the derived column densities and excitation temperatures. If the beam dilution effect is included in the rotation diagram, it is noticed that, on average, our column densities and temperatures would increase by a factor of  $3.2$ and $1.4$, respectively.
 Derived column density and rotational temperature are further used in the LTE module of CASSIS to produce synthetic LTE spectra and also compare with our observations (see Figure \ref{fig:fit1}). The average FWHM of each species from Table \ref{tab:dataobs} is used for this computation. A source size of 2$^{''}$ with the ALMA telescope is considered. Figure \ref{fig:fit1} shows a good agreement with the observed feature except for the three transitions of methanol and all the transitions of dimethyl ether. The intensity of the three methanol transitions is overproduced. Among them, $153.281$ GHz and $154.425$ GHz transitions are optically thick.  All the transitions of dimethyl ether are underproduced. 
During the line analysis, it is noticed that these transitions consist of multiple sub-states. The strongest transition is only considered based on their S$\mu^2$ value. Thus, the obtained FWHMs are possibly overestimated during the Gaussian fitting, which can under-produce the intensity of these transitions in Figure \ref{fig:fit1}.


\begin{deluxetable*}{ccccccccc}
\tablecaption{ Calculated Gibbs free energy of activation, activation barrier, and reaction enthalpy of alochols formation via successive hydrogen addition with aldehydes, ketone, and ketene by using lower-level [B3LYP/6-31+G(d,p)] and higher-level [CCSD(T)/aug-cc-pVTZ] of theory considering $298.15$ K temperature and $1$ atmospheric pressure.
\label{tab:reac}}
\tablewidth{0pt}
\tabletypesize{\scriptsize} 
\tablehead{
\colhead{Serial No.} & \colhead{Reaction type} & \colhead{Reactions} & \multicolumn{2}{c}{Gibbs Free Energy of Activation} & \multicolumn{2}{c}{Activation Barrier} & \multicolumn{2}{c}{Enthalpy of Reaction} \\
\colhead{ } & \colhead{ } & \colhead{ } & \multicolumn{2}{c}{(kcal/mol)} & \multicolumn{2}{c}{(kcal/mol)} & \multicolumn{2}{c}{(kcal/mol)} \\
\cline{4-9}
\colhead{ } & \colhead{ } & \colhead{ } & \colhead{Gas Phase} & \colhead{Ice Phase} & \colhead{Gas Phase} & \colhead{Ice Phase} & \colhead{Gas Phase} & \colhead{Ice Phase}
}
\startdata
&&&&&&\\
\multicolumn{9}{c}{Methanal - Methanol} \\
&&&&&&\\
1a. &(NR)& H$_2$CO + H $\rightarrow$ CH$_2$OH & 10.57 & 11.32 & 5.59 (9.35)$^a$[10.35]$^i$ & 6.49 (10.02)$^a$[11.16]$^i$ & -32.31 & -32.39 \\
1b. &(RR)& CH$_2$OH + H $\rightarrow$ CH$_3$OH & \nodata & \nodata & 0.00  & 0.00  & -94.75 & -94.93 \\
2a. &(NR)& H$_2$CO + H $\rightarrow$ CH$_3$O & 7.10 & 6.97 & 1.89 (3.35)$^a$[4.29]$^i$& 1.76 (3.10)$^a$[3.26]$^i$& -27.61 & -27.07 \\
2b. &(RR)& CH$_3$O + H $\rightarrow$ CH$_3$OH & \nodata & \nodata & 0.00 & 0.00 & -99.44 & -100.24 \\
\hline
&&&&&&\\
\multicolumn{9}{c}{Ethanal - Ethanol} \\
&&&&&&\\
3a. &(NR)& CH$_3$CHO + H $\rightarrow$ CH$_3$CHOH & 11.41 & 12.44 & 5.91 (9.38)$^a$ & 6.91 (10.24)$^a$ & -27.96 & -26.80 \\
3b. &(RR)& CH$_3$CHOH + H $\rightarrow$ CH$_3$CH$_2$OH & \nodata & \nodata & 0.00 & 0.00 & -92.61 & -93.21 \\
4a. &(NR)& CH$_3$CHO + H $\rightarrow$ CH$_3$CH$_2$O & 10.10 & 10.24 & 4.11 (5.36)$^a$& 4.25 (5.38)$^a$& -21.50 & -20.27 \\
4b. &(RR)& CH$_3$CH$_2$O + H $\rightarrow$ CH$_3$CH$_2$OH & \nodata & \nodata & 0.00 & 0.00 & -99.07 & -99.74 \\
\hline
&&&&&&\\
\multicolumn{9}{c}{Propanal - 1-Propanol} \\
&&&&&&\\
5a. &(NR)& CH$_3$CH$_2$CHO + H $\rightarrow$ CH$_3$CH$_2$CHOH & 11.12 & 12.08 & 5.47 (9.27)$^a$[9.82]$^{ii}$& 6.41 (9.87)$^a$& -28.10 [-26.31]$^{ii}$ & -18.42 \\
5b. &(RR)& CH$_3$CH$_2$CHOH + H $\rightarrow$ CH$_3$CH$_2$CH$_2$OH & \nodata & \nodata & 0.00 & 0.00 & -92.82 & -101.80 \\
6a. &(NR)& CH$_3$CH$_2$CHO + H $\rightarrow$ CH$_3$CH$_2$CH$_2$O & 10.08 & 10.27 & 3.99 (4.97)$^a$[5.71]$^{ii}$& 4.19 (4.96)$^a$ & -22.00 [-19.74]$^{ii}$   & -20.76 \\
6b. &(RR)& CH$_3$CH$_2$CH$_2$O + H $\rightarrow$ CH$_3$CH$_2$CH$_2$OH & \nodata & \nodata & 0.00 & 0.00 & -98.91 & -99.46 \\
\hline
&&&&&&\\
\multicolumn{9}{c}{Propenal - Allyl alcohol} \\
&&&&&&\\
7a. &(NR)& CH$_2$CHCHO + H $\rightarrow$ CH$_2$CHCHOH & 9.56 & 10.48 & 4.00(6.30)$^a$[7.89]$^{ii}$& 2.36 (10.47)$^a$ & -40.29 [-38.93]$^{ii}$&  -39.59\\
7b. &(RR)& CH$_2$CHCHOH + H $\rightarrow$ CH$_2$CHCH$_2$OH & \nodata & \nodata & 0.00 & 0.00 & -76.72 & -77.07 \\
8a. &(NR)& CH$_2$CHCHO + H $\rightarrow$ CH$_2$CHCH$_2$O & 10.32 & 10.24 & 4.43 (5.79)$^a$[6.43]$^{ii}$&4.38 (6.60)$^a$ &-18.00 [-16.16]$^{ii}$& -16.72\\
8b. &(RR)& CH$_2$CHCH$_2$O + H $\rightarrow$ CH$_2$CHCH$_2$OH & \nodata & \nodata & 0.00 & 0.00 & -99.00 & -99.95 \\
\hline
&&&&&&\\
\multicolumn{9}{c}{Propynal - Propargyl alcohol} \\
&&&&&&\\
9a. &(NR)& HC$_2$CHO + H $\rightarrow$ HC$_2$CHOH & 8.86 & 9.65 & 3.41 [7.41]$^{ii}$ & 4.20 (8.28)$^a$ & -43.92 [-42.42]$^{ii}$ & -43.22 \\
9b. &(RR)& HC$_2$CHOH + H $\rightarrow$ HC$_2$CH$_2$OH & \nodata & \nodata & 0.00 & 0.00 & -79.53 & -80.44 \\
10a. &(NR)& HC$_2$CHO + H $\rightarrow$ HC$_2$CH$_2$O & 9.63 & 9.47 & 3.74 [5.78]$^{ii}$ & 3.59 (5.68)$^a$ & -22.42 [-21.00]$^{ii}$ & -22.20 \\
10b. &(RR)& HC$_2$CH$_2$O + H $\rightarrow$ HC$_2$CH$_2$OH & \nodata & \nodata & 0.00 & 0.00 & -101.02 & -101.46 \\
\hline
&&&&&&\\
\multicolumn{9}{c}{Glycolaldehyde - Ethylene glycol} \\
&&&&&&\\
11a. &(NR)& HOCH$_2$CHO + H $\rightarrow$ HOCH$_2$CHOH & 11.42 & 12.31 & 5.50 (8.75)$^a$[9.51]$^{iii}$& 6.32 (9.29)$^a$& -29.43 & -30.43 \\
11b.&(RR) & HOCH$_2$CHOH + H $\rightarrow$ HOCH$_2$CH$_2$OH & \nodata & \nodata & 0.00 & 0.00 & -92.18 & -92.44 \\
12a. &(NR)& HOCH$_2$CHO + H $\rightarrow$ HOCH$_2$CH$_2$O & 9.43 & 9.67 & 3.34 (4.47)[4.97]$^{iii}$& 3.38 (4.17)$^a$ & -22.38 & -22.61 \\
12b. &(RR)& HOCH$_2$CH$_2$O + H $\rightarrow$ HOCH$_2$CH$_2$OH & \nodata & \nodata & 0.00 & 0.00 & -99.22 & -100.25 \\
\hline
&&&&&&\\
\multicolumn{9}{c}{Acetone - Isopropanol} \\
&&&&&&\\
13a. &(NR)& CH$_3$COCH$_3$ + H $\rightarrow$ CH$_3$COHCH$_3$ & 11.27 & 12.44 & 5.56 (9.15)$^a$&6.66 (9.85)$^a$ & -25.76 & -24.35 \\
13b. &(RR)& CH$_3$COHCH$_3$ + H $\rightarrow$ CH$_3$CHOHCH$_3$ & \nodata & \nodata & 0.00 & 0.00 & -90.73 & -91.08 \\
14a. &(NR)& CH$_3$COCH$_3$ + H $\rightarrow$ CH$_3$CHOCH$_3$ & 12.97 & 12.97 & 6.29&6.55 (---)$^b$& -16.40 & -24.35 \\
14b. &(RR)& CH$_3$CHOCH$_3$ + H $\rightarrow$ CH$_3$CHOHCH$_3$ & \nodata & \nodata & 0.00 & 0.00 & -100.10 & -91.08 \\
\hline
&&&&&&\\
\multicolumn{9}{c}{Ethenone - Vinyl alcohol} \\
&&&&&&\\
15a. &(NR)& CH$_2$CO + H $\rightarrow$ CH$_2$COH & 17.52 & 18.22 &11.80 (5.39)&12.49 (17.59)$^a$ & -15.60  & -16.96 \\
15b. &(RR)& CH$_2$COH + H $\rightarrow$ CH$_2$CHOH & \nodata & \nodata & 0.00 & 0.00 & -45.67 & -45.30 \\
16a. &(NR)& CH$_2$CO + H $\rightarrow$ CH$_2$CHO & 10.04 & 9.92 &4.36 (3.97)&4.26 (5.43)$^a$ & -41.76 & -43.87 \\
16b. &(RR)& CH$_2$CHO + H $\rightarrow$ CH$_2$CHOH & \nodata & \nodata & 0.00 & 0.00 & -19.51 & -18.40 \\
\enddata
\tablecomments{
 NR refers to Neutral-Radical reactions, and RR to Radical-Radical reactions.\\ 
$^a$Bracketed values are calculated using CCSD(T)/aug-cc-pVTZ level, $^b$We did not find true transition state, $^i$\cite{song17}, $^{ii}$\cite{zave18}, $^{iii}$\cite{barc18}.}
\end{deluxetable*}

\begin{figure*}
\centering
\includegraphics[width=0.9\textwidth]{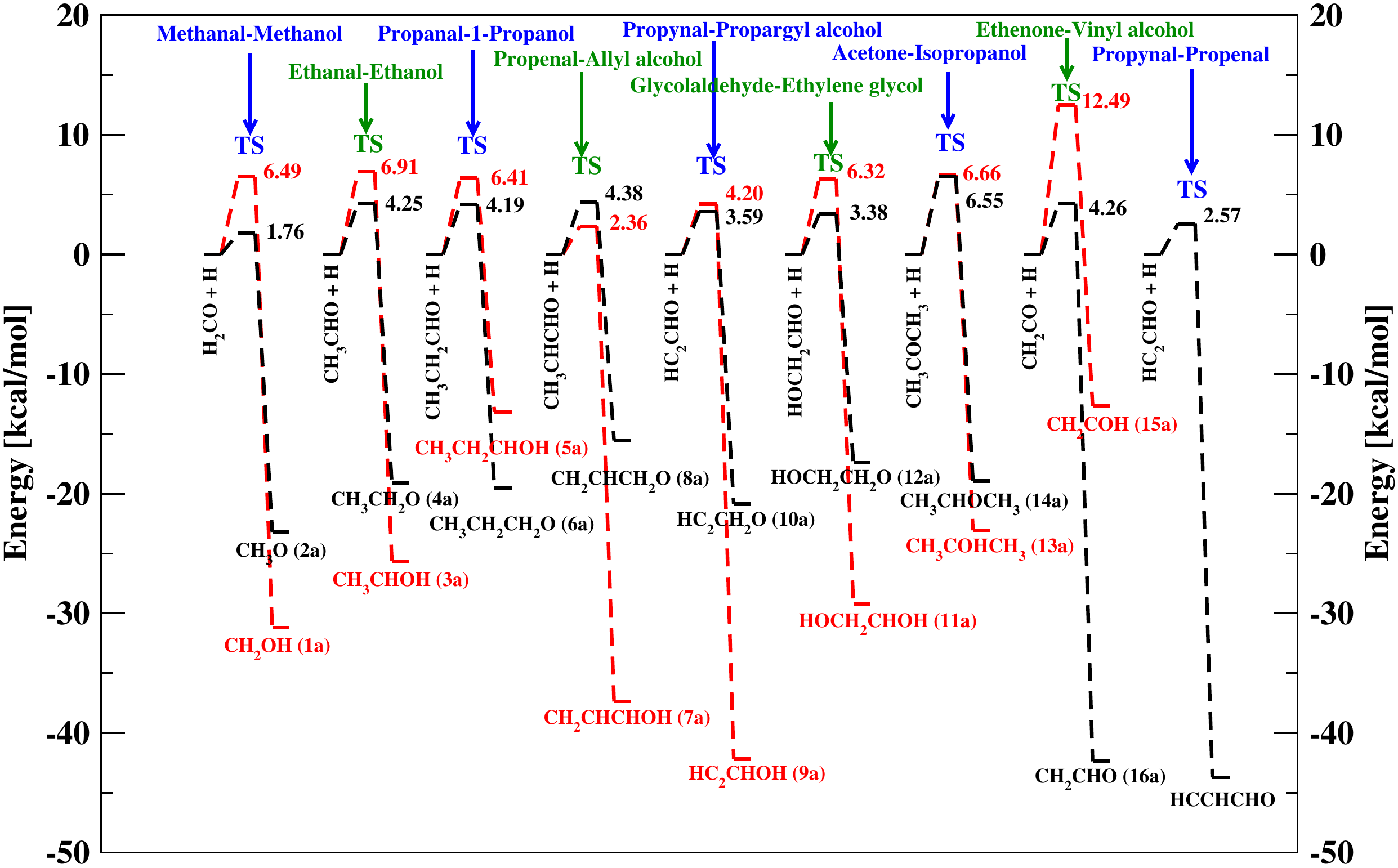}
\caption{Schematic potential energy surfaces of alcohol formation from aldehyde and ketone for various hydrogenation reactions (representing IEFPCM correction for the influence of ice) indicating stable species 
and transition states. Energy differences in units of kcal/mol are shown, referenced to reactant complexes calculated using DFT-B3LYP/6-31G+(d,p) level of theory. The red lines denote H addition at O positions, whereas the black lines represent H addition at C positions.}
\label{fig:TS}
\end{figure*}

\section{Kinetics of the aldehydes, alcohols, ketone, and ketene observed in G10} \label{sec:kinetics}
Here, an extensive theoretical investigation is carried out to understand the formation mechanism of some aldehydes, alcohols, ketene, and ketone, which are observed in G10. 
All the quantum chemical results presented in this article are carried out using the Gaussian 09 suite of programs \citep{fris13}. The Density Functional Theory (DFT) method with B3LYP functional \citep{beck93} and 6-31+G(d,p) basis set is used in computing the activation barrier and enthalpy of reactions. Both the activation and reaction energies are calculated including the harmonic zero-point vibrational energy (ZPVE) and considering a temperature of $\sim 298.15$ K and $\sim 1$ atmosphere pressure. A fully optimized ground-state structure is verified to be a stationary point (having non-imaginary frequency) by harmonic
vibrational frequency analysis. For the transition state (TS) calculation, the QST2 method of the Gaussian 09 program is used. All the TSs
are verified using the internal reaction coordinate (IRC) analysis which connects reactants and products through the minimum energy path of the potential energy surface (PES). Further, the higher-order method is employed to examine the effect on the activation barrier calculation. The single point energy calculation of the TS structure and the optimized structure of the reactants at
CCSD(T)/aug-cc-pVTZ level is carried out. The ice mixture is considered embedded within a continuum solvation field to include the passive impact of bulk ice on the activation barrier and enthalpy of reaction. 
To mimic the ice phase feature,  water is used as the solvent. A self-consistent reaction field (SCRF) is routinely applied in the electronic structure calculations to study the spectroscopy and thermochemistry parameters of any species in the ice phase. Explicit water molecules can also generate a more realistic (inhomogeneous) field. Since the typical size of the interstellar water cluster (numbers of water molecules required) is not known with certainty, it is preferred to use the SCRF method \citep{woon02b}.
The Polarizable Continuum Model (PCM) with different integral equation formalism variant (IEFPCM) as a default self-consistent reaction field (SCRF) method is used for this purpose \citep{canc97,toma05}. A similar level of theory and basis set is used for the gas phase calculations. This method creates a solute cavity by overlapping spheres \citep{toma94,pasc87}. The implicit solvation
model places the molecule of interest inside a cavity in a continuous homogeneous dielectric medium (in this case, water) that represents the solvent. This method is proven to be producing convincing results for ices. The PCM model is used to consider the reaction field of bulk ice. Here, the cluster is treated explicitly to avoid spurious boundary effects. In this framework, the reaction fields issuing from a dielectric constant of $78$ for water or about 100 for ice \citep{arag11} are essentially identical. Hence, the solid phase calculations and the thermochemistry parameters reported here are close to those in ice.
The BE and PA of the aldehydes, ketone, ketene, and corresponding alcohols are also calculated quantum chemically discussed in Section \ref{sec:BE_PA}.

\subsection{Rate constants}
A cross-over temperature, T$_c$ is defined as the maximum temperature below which tunneling reaction dominates and above which thermal activation reaction takes over. T$_c$ is calculated by the following relation:
\begin{equation}
 T_c=\frac{h\omega}{4\pi^{2} k_B},
\end{equation}
where h is the Planck's constant, k$_B$ is the Boltzmann constant, and $\omega$ is the absolute value of the 
the imaginary frequency at the TS. 
For all the gas phase reactions, the transition state theory (TST) is applied. 
Gas-phase rate constant can be calculated by the following expression,
\begin{equation}
K=\Big(\frac{k_BT}{hc^{0}}\Big)\exp(-\Delta {G}^{\ddag}/RT),
\end{equation}
where, $\rm{\Delta {G}^{\ddag}}$ is the Gibbs free energy of activation, $c^{0}$ is the concentration ($\sim 1$), $R$ is the ideal gas constant and $T$ is the temperature.

High-level quantum chemical calculations of six aldehyde-alcohol pairs with one ketone-alcohol and one ketene-alcohol 
pairs are presented and
discussed in detail. Aldehyde to alcohol, ketone to alcohol, ketene to alcohol formation may occur 
through successive hydrogen addition reaction in two
steps: The first step is aldehyde/ketone/ketene + H $\rightarrow$ radical-1 and the second step is radical-1 + H $\rightarrow$ alcohol. The first step of all these reactions is neutral-radical reactions, whereas the second step is a radical-radical reaction. The first step of all these reactions thus having an activation barrier, whereas the second step is barrierless as both reactants have free electrons, which makes them very reactive. 
Our calculations show that six aldehydes (such as methanal, ethanal, propanal, propenal, propynal, and glycolaldehyde), one ketone (such as acetone), and one ketene (such as ethenone) are forming alcohols (such as methanol,
ethanol, 1-propanol, allyl alcohol, propargyl alcohol, ethylene glycol, isopropanol, and vinyl alcohol,
respectively) via two successive hydrogen addition reactions. 
In terms of stability, the observed alcohols are thermodynamically favorable compared to their isomers. 
The formation of complex molecules on the grain surface through successive hydrogen additions is found to be dominant.
Formation pathways of these molecules are discussed below, and a graphical representation of PESs of their formation mechanism considering IEFPCM model and DFT-B3LYP/6-31G+(d,p) level of theory is shown in Figure \ref{fig:TS}.

\subsubsection{Methanal and Methanol}
Here, the activation barrier of the reaction H +H$_2$CO is computed, for the production of $\rm{CH_2OH}$ or $\rm{CH_3O}$ (reactions 1a and 2a of Table \ref{tab:reac}). It is noticed that the hydrogen addition to the carbon atom of H$_2$CO (i.e., the formation of CH$_3$O, methoxy radical) is more
favorable than that of its addition to the O atom of H$_2$CO (i.e., the formation of CH$_2$OH) in both gas- and ice-phases considering both (DFT-B3LYP/6-31+G(d,p) and CCSD(T)/aug-cc-pVTZ) level of theory.
This is consistent with earlier experimental and theoretical results \citep{chua16,song17}.
\cite{song17} studied the TS of reactions 1a and 2a both in gas-phase and on five different binding sites of the amorphous solid water (ASW) surface and calculated the corresponding activation energies. We compare our activation energy values with their gas-phase and ice-phase for binding site ASW 5 in Table \ref{tab:reac}, which shows a good agreement when we considered CCSD(T)/aug-cc-pVTZ level of theory.
\cite{woon02a} predicted the barrier height of reaction 2a as 5.04 kcal/mol for the gas-phase and it decreases to 4.81 kcal/mol when the reaction is embedded with the IPCM field. In our case, we also found the similar decreasing trend from gas-phase to ice-phase considering both level of theory.
\cite{song17} also computed the BE of H$_2$CO for different sites of amorphous solid water (ASW) surface and found that BE values vary between $1000$ K and $9370$ K. This uncertainty in the BE value may lead to a very different abundance of CH$_3$OH
under different astrophysical environments.
The hydrogen abstraction reaction of methanol also plays an essential role in controlling the abundance of COMs. Methanol
abstraction reaction barrier and the rate constant are taken from \cite{song17}.

\subsubsection{Ethanal and Ethanol}
 Here, the formation of
$\rm{C_2H_5OH}$ via two successive hydrogen addition reactions with the CH$_3$CHO is considered. Kinetic information (i.e., activation barrier and reaction enthalpy) of these two reactions are noted in Table \ref{tab:reac}.
Looking at the activation barriers reported in Table \ref{tab:reac}, it is clear
that similar to methanol formation, here also, the hydrogen
addition at the carbon atom position of CH$_3$CHO (i.e., the formation of CH$_3$CH$_2$O, reaction 4a) is more efficient compared
to the hydrogen addition at the O atom position (i.e., the formation of CH$_3$CHOH, reaction 3a). 
The relative abundances between the acetaldehyde and
ethyl alcohol seem  to depend upon the evolution time, availability of the hydrogen atoms, ice thickness, and ice
morphology \citep{biss07,biss08}. 
Only the hydrogenation reactions are not enough in explaining the observed abundances of ethanal-ethanol. It is needed to 
consider the ice phase reaction between CH$_3$ and CH$_2$OH for the formation of ethanol in the ice phase.
\subsubsection{Propanal and 1-Propanol}
The coexistence of propanal, propynal, and propenal in Sgr B2(N) \citep{holl04} could be explained by the following successive hydrogenation sequences;
$\rm{HC_2CHO\ (propynal) \ \rightarrow \ CH_2CHCHO\ (propenal)}$  $\rm{\rightarrow \ CH_3CH_2CHO\ (propanal)}$.
Recently, \cite{qasi19} studied the propanal formation by the radical-radical reaction
between HCO and $\rm{H_3CCH_2}$. They used the sequential hydrogenation of propanal to form
1-propanol. The first step of the reaction has a high activation barrier. But this reaction can change the
saturated bond to the unsaturated radicals, which may again undergo some
radical-radical recombination reactions.
\cite{qasi19} obtained a significant production of 1-propanol with this pathway. Our TST calculation for the successive
hydrogenation pathways is shown in Table \ref{tab:reac}. It is noticed that the
hydrogen addition at the carbon atom position of propanal ($\rm{CH_3CH_2CH_2O}$ formation, reaction 6a of Table \ref{tab:reac}) is favorable compared to the oxygen atom position of propanal ($\rm{CH_3CH_2CHOH}$ formation, reaction 5a). The gas-phase activation and reaction energies (including ZPVE) of \cite{zave18} are compared with ours in Table \ref{tab:reac} which is in good agreement with our values calculated using higher-level of theory. Except the successive hydrogenation reactions, following ice phase reactions, are also considered for the formation of propynal, propenal, and propanal.
$$
{\rm C_2H + H_2CO \rightarrow propynal +H},
$$
$$
{\rm O + C_3H_3 \rightarrow propynal +H},
$$
$$
{\rm HCO + C_3H_3 \rightarrow propenal +H},
$$
$$
{\rm HCO + H_3CCH_2 \rightarrow propanal +H}.
$$

\subsubsection{Propenal and Allyl Alcohol}
Propenal is also a product of the two successive hydrogenation reactions with propynal.
$$
{\rm propynal + 2H \rightarrow propenal \ (CH_2CHCHO). }
$$
 Here, the other product channel of this reaction is studied. Firstly, the hydrogenation reaction of propynal would form propenal. In the second step, propenal could undergo another hydrogenation reaction to form allyl alcohol. 
The TS calculation of the first step found an activation barrier of 2.60 kcal/mol in the gas phase and 2.57 kcal/mol in the ice phase considering lower-level of theory (see Figure \ref{fig:TS}). Table \ref{tab:reac} shows the activation barriers for the hydrogen addition at the propenal's carbon and oxygen atom position. It is noted that, unlike the other cases, here, the hydrogen addition at the oxygen atom of propenal ($\rm{CH_2CHCHOH}$ formation, reaction 7a) seems to be more favorable than it is with the carbon atom position ($\rm{CH_2CHCH_2O}$ formation, reaction 8a)  considering lower-level of theory. But considering higher-level of theory, opposite trend is found which agrees well with the values calculated by \cite{zave18} shown in Table \ref{tab:reac} for comparison.

\subsubsection{Propynal and Propargyl Alcohol}
Transformation from  propynal ($\rm{HC_2CHO}$) to propargyl alcohol was already reported by \cite{gora17b}. This happened via two successive hydrogenations with propynal :
$$
{\rm propynal + 2H \rightarrow propynol \ (propargyl \ alcohol, \ HC_2CH_2OH). }
$$
Table \ref{tab:reac} shows 
a different trend for the hydrogen addition for reactions 9a and 10a in the gas-phase. The hydrogen addition to the oxygen atom position (reaction 9a) is favorable than that of the carbon atom position (reaction 10a) using lower-level of theory, whereas \cite{zave18} found an opposite trend. In contrast, the opposite trend is noticed for the ice-phase using both lower- and higher-level of theory.

\subsubsection{Glycolaldehyde and Ethylene Glycol}
Glycolaldehyde is the simplest form of the aldose family. The sugars like glucose, ribose, and erythrose belong to this family.
Therefore, the presence of glycolaldehyde in space is an indication of the precursor of biologically relevant molecules. Identification of the numerous complex organic molecules in the Galactic Centre (GC) makes it an active area of research.  Specifically, the study of the
chemical evolution of the grain mantle, where various energetic processes (like UV radiation, X-rays, and cosmic rays, etc.) may
reprocess the mantle composition, are fascinating. The C and O addition to the formyl radical and then three successive hydrogen additions
yield glycolaldehyde. The two subsequent
hydrogen additions to the glycolaldehyde can produce ethylene glycol \citep{char01}.
\cite{fedo15a} proposed that two HCO radicals can recombine to produce glyoxal (HC(O)CHO), which can also subsequently convert to glycolaldehyde. \cite{cout18} investigated the
formation mechanism of these two species (glycolaldehyde and ethylene glycol) using a gas-grain
chemical model. They studied the EG/GA ratio for
various luminosities. 
\cite{barc18} studied two successive hydrogen addition reactions with acetaldehyde which can produce ethylene glycol.
They also studied the hydrogen abstraction reactions of GA and EG. Table \ref{tab:reac} shows the kinetics involved in converting ethylene glycol from glycolaldehyde. Like the methanal-methanol, ethanal-ethanol, propanal-1-propanol, and ice phase pathways of propynal-propargyl alcohol pairs, the favorable position of the hydrogen addition is noted to be the carbon atom ($\rm{HOCH_2CH_2O}$ formation, reaction 12a) instead of the oxygen atom ($\rm{HOCH_2CHOH}$ formation, reaction 11a) of glycolaldehyde.

\subsubsection{Acetone and Isopropanol}
Acetone ($\rm{CH_3COCH_3}$, propanone) and isopropanol ($\rm{CH_3CHOHCH_3}$, isopropyl alcohol, 2-propanol) might be chemically connected.
Table \ref{tab:reac} shows the
two successive hydrogenations of acetone (reactions 13 and 14). 
We check the TST calculations for the hydrogen
addition either with the oxygen atom (activation energy 5.56 kcal/mol) or with the carbon atom (activation energy 6.29 kcal/mol) position of acetone and found that the former one is more favorable than the latter in the gas phase considering lower-level of theory. In contrast, the opposite trend was noted in the ice phase. We could not find a proper TS using higher-level of theory for reaction 14a in the ice-phase.

\subsubsection{Ethenone and Vinyl Alcohol}
 Vinyl alcohol is a planar molecule, which has two conformations. The ``syn" conformer is found to be more stable than that of the ``anti"  conformer \citep{turn01}. Here also,  the ``syn" conformer of vinyl alcohol is considered for the quantum chemical calculations.
Here, the TS of the two successive hydrogen addition reactions to ketene is examined for the formation of vinyl alcohol.
The first step has an activation barrier, whereas the second step is a radical-radical reaction and is assumed to be barrierless.
Vinyl alcohol can also form from acetaldehyde via the isomerization process \citep{kase20}. Table \ref{tab:reac} depicts that the hydrogen addition at the carbon atom position is more favorable than the oxygen atom position of ethenone. 
Table \ref{tab:reac} shows that the H addition to the O atom position of ketene (reaction 15a) requires a significantly higher activation barrier than the H addition to the O atom position of the other species (reactions 1a, 3a, 5a, 7a, 9a, 11a, 13a) reported here. It could be due to the change of the hybridization of C atom of ketene from sp (linear) to sp2 (bent) during the H addition reaction. On the other hand, there is no change of hybridization upon H addition to the other aldehyde, which requires only small structural change and thus, the activation energy could be small. \cite{carr68} studied the reaction of ketene with atomic H at room temperature where the reaction $\rm{H+ H_2CCO \rightarrow CH_3CO^* \rightarrow CH_3 + CO}$ occurs rather slowly (rate constant $1.3\times10^{-13}$ cm$^3$mol$^{-1}$s$^{-1}$), while neither $\rm{H + H_2CCO \rightarrow CH_2CHO}$ nor H abstraction occur very significantly.

\begin{deluxetable*}{lcccccc}
\tablecaption{Calculated binding energy values of the aldehydes, alcohols, ketone and ketene considered in this work.\label{tab:substrate}}
\tablewidth{0pt}
\tabletypesize{\scriptsize} 
\tablehead{
\colhead{Species} & \colhead{Ground state} & \multicolumn{5}{c}{Binding Energy (Kelvin)} \\
\cline{3-7}
\colhead{ } & \colhead{used} & \colhead{CO monomer} & \colhead{CH$_3$OH monomer} & \colhead{H$_2$O monomer} & \colhead{H$_2$O tetramer (scaled by 1.188)} & \colhead{Available}
}
\startdata
Methanal  & Singlet & 641 & 2942 & 2896 & 3242$^a$ (3851) & 2050$^b$, 4500 $\pm$ 1350$^c$ \\
Methanol  & Singlet & 1247 & 3227 & 3124 & 4368$^a$ (5189) & 5534$^b$, 5000 $\pm$ 1500$^c$ \\
\hline
Ethanal & Singlet & 1015& 1964 &3189& 3849$^a$ (4573) & 2450$^b$, 5400 $\pm$ 1620$^c$ \\
Ethanol  & Singlet &873& 3408 &2824& 5045 (5993) & 6584$^b$, 5400 $\pm$ 1620$^c$ \\
\hline
Propanal  & Singlet & 680 & 3679 &3522& 3396 (4034) & 4500 $\pm$ 1350$^c$ \\
1-Propanol  & Singlet & 1298 & 3932 &2609& 4791 (5692) & \nodata \\
\hline
Propenal & Singlet & 1043 & 3236 &3170& 3495 (4152) & 5400 $\pm$ 1620$^c$ \\
Allyl Alcohol  & Singlet & 1381 & 3625 &2958& 5964 (7085) & \nodata \\
\hline
Propynal &Singlet&1174&3072&2318& 3379 (4014) & \nodata \\
Propargyl Alcohol&Singlet& 1031 & 4119 & 2870 & 4284 (5089) & \nodata \\
\hline
Glycolaldehyde  & Singlet & 826 & 3254 & 2871 & 5131 (6096) & \nodata \\
Ethylene Glycol  & Singlet & 933 & 4666 & 3144 & 4602 (5467) & \nodata \\
\hline
Acetone & Singlet & 728 &3769&3623& 4420 (5251) & 3500$^b$ \\
Isopropanol & Singlet & 1369 &3640 &2879& 5230 (6213) & \nodata \\
\hline
Ethenone & Singlet & 732 & 1450 & 1317 & 2847 (3382) & 2200$^b$, 2800 $\pm$ 840$^c$ \\
Vinyl Alcohol & Singlet & 1236 & 3407 & 2836 & 3786 (4497) & \nodata \\
\enddata
\tablecomments{
$^a$\cite{das18}, $^b$KIDA old BE, $^c$KIDA new BE \citep{wake17}.
}
\end{deluxetable*}

\begin{deluxetable*}{lccc}
\tablecaption{Calculated proton affinity of these species.
\label{tab:pa}}
\tablewidth{0pt}
\tabletypesize{\scriptsize} 
\tablehead{
\colhead{Species} & \colhead{Protonation Reactions} & \multicolumn{2}{c}{Proton Affinity (kJ/mol)} \\
\cline{3-4}
\colhead{ } & \colhead{ } & \colhead{Our calculation} & \colhead{Available} 
}
\startdata
 Methanal & HCHO + H$^+$ $\rightarrow$ CH$_2$OH$^+$ & 692  & 712.9 $\pm$ 1.1$^a$  \\
 Methanol & CH$_3$OH + H$^+$ $\rightarrow$ CH$_3$OH$_2^+$ & 740 & 754.3$^a$ \\
\hline
 Ethanal & CH$_3$CHO + H$^+$ $\rightarrow$ CH$_3$CHOH$^+$ & 749  & 768.5 $\pm$ 1.6$^a$ \\
 Ethanol & C$_2$H$_5$OH + H$^+$ $\rightarrow$ C$_2$H$_5$OH$_2^+$  & 764 &  \\
\hline
 Propanal & CH$_3$CH$_2$CHO + H$^+$ $\rightarrow$ HCH$_3$CH$_2$CHO$^+$ & 769 &  \\
 1-Propanol & CH$_3$CH$_2$CH$_2$OH + H$^+$ $\rightarrow$ HCH$_3$CH$_2$CH$_2$OH$^+$ & 770 & \\
\hline
 Propenal & CH$_2$CHCHO + H$^+$ $\rightarrow$ CH$_2$CHCHOH$^+$ & 764 & \\
 Allyl Alcohol & CH$_2$CHCH$_2$OH + H$^+$ $\rightarrow$ CH$_2$CHCH$_2$OH$_2^+$ & 773 & \\
 \hline
Propynal & $\rm{HC_2CHO + H^+ \rightarrow HC_2CHOH^+}$ & 733 &  \\
Propargyl Alcohol & $\rm{HC_2CH_2OH + H^+ \rightarrow HC_2CH_2OH_2^+}$ & 742 & \\
 \hline
 Glycolaldehyde & HOCH$_2$CHO + H$^+$ $\rightarrow$ H$_2$OCH$_2$CHO$^+$ & 759 & \\
 Ethylene Glycol & HOCH$_2$CH$_2$OH + H$^+$ $\rightarrow$ HOCH$_2$CH$_2$OH$_2^+$ & 797 & \\
\hline
 Acetone & CH$_3$COCH$_3$ + H$^+$ $\rightarrow$ CH$_3$COHCH$_3^+$ & 793 &  \\
 Isopropanol & CH$_3$CHOHCH$_3$ + H$^+$ $\rightarrow$ CH$_3$CHOH$_2$CH$_3^+$ & 782  & \\
\hline
 Ethenone & CH$_2$CO + H$^+$ $\rightarrow$ CH$_3$CO$^+$ & 822 & 825.3$\pm$3$^a$ \\
 Vinyl Alcohol & CH$_2$CHOH + H$^+$ $\rightarrow$ CH$_3$CHOH$^+$ & 800 & 882.8$^b$ \\
\enddata
\tablecomments{$^a$\cite{hunt98}, $^b$\cite{turn01}}
\end{deluxetable*}

\begin{deluxetable*}{cccccccc}
\tablecaption{Comparison between the observed abundance and our CMMC results with Model C (set 4).
\label{tab:comp}}
\tablewidth{0pt}
\tabletypesize{\scriptsize} 
\tablehead{
\colhead{Species} & \colhead{Obtained abundances} & \colhead{Obtained abundance} & \multicolumn{4}{c}{ Observed abundances}\\
\cline{4-7}
\colhead{ } & \colhead{from model } & \colhead{ratio from model}& \colhead{G31.41+0.31} & \colhead{Sgr B2 (N)} & \colhead{Orion KL } & \colhead{G10.47+0.03 }
}
\startdata
Methanal& $3.3(-9)-5.8(-7)$&&$[1.9(-7)]^e$&$[3.2(-08)]^m$&&$[6.1(-7)]^b$ \\
Methanol& $3.5(-8)-1.4(-5)$ & $1.0(1)-5.3(1)$&$[1.9(-6)]^e$& $[7.7(-7)]^m$ $[7.1(-6)]^x$&$[2.2(-7)]^l$ & $[3.2\pm2.5 (-7)]^a$, $[1.8(-6)]^b$ \\
\hline
Ethanal&$1.8(-13)-5.0(-9)$&&$[1.0(-9)]^n$&$[1.4(-08)]^m$ $[9.5(-8)]^x$&&$[5.0\pm0.8(-10)]^a$ \\
Ethanol & $7.4(-10)-9.7(-8)$ & $8.0(-2)-2.6(4)$ & $[2.0(-8)]^n[1.3(-9)]^f$ & $[1.1(-7)]^m [3.6(-7)]^x$ & $[1.3(-8)]^l$ & $[1.2(-7)]^b$ \\
\hline
Propanal&$7.3(-10)-4.8(-8)$ &&&&& $[1.8\pm0.2(-8)]^a$ \\
1-Propanol& $2.4(-12)-1.1(-7)$ & $2.6(-4)-1.3(2)$ &&&&\\
\hline
Propenal&$2.0(-13)-1.9(-9)$&&&&&$[1.2(-10)]^{*d}$ \\
Allyl alcohol& $8.7(-15)-1.7(-9)$ & $3.6(-2)-2.9(0)$ &&&&\\
\hline
Propynal & $3.1(-11)-2.2(-09)$ &&&&&\\
Propargyl alcohol & $6.4(-15)-1.3(-11)$ & $1.4(-4)-1.0(-2)$ &&&&\\
\hline
Glycolaldehyde & $2.8(-15)-1.3(-06)$ && $[8.3(-11)]^f$& $[3.2(-10)]^m$ $[1.9(-8)]^x$& $[1.5(-10)]^l$ & [$9.6\pm0.007(-10)]^a$ \\
Ethylene glycol & $5.5(-17)-1.6(-5)$ & $1.9(-2)-4.45(2)$ &$[1.7(-10)]^f$&$4.1(-10)]^m$ & $[2.3(-9)]^l$ &$[3.7\pm1.5(-8)]^a$ \\
\hline
Acetone& $6.6(-14)-1.7(-8)$ &&& $[2.6(-8)]^m$ &&$[4.0\pm0.5(-8)]^a,\ [1.0(-7)]^b$\\
Isopropanol& $1.2(-13)- 4.4(-9)$& $9.7(-2)-7.4(3)$&&&&\\
\hline
Ethenone& $4.6(-13)-5.5(-9)$ &&&&&$[8.0(-8)]^b$ \\
Vinyl alcohol & $3.5(-11)-1.4(-8)$ &$1.9(0)-7.7(1)$ &&&&$[{\leq} 2.3(-9)]^{*d}$ \\
\hline
Methyl formate&$2.3(-11)-1.3(-6)$ && $[2.6(-9)]^f$&$[7.7(-8)]^m$ $[9.5(-8)]^x$ &$[1.2(-7)]^k$ &$[6.7\pm0.5(-8)]^a,\ [1.4(-7)]^b$ \\
Dimethyl ether&$3.5(-9)-1.5(-7)$ & $1.2(-1)-1.5(2)$ & $[4.2(-9)]^f$ & $[3.5(-7)]^m$ &$[3.0(-7)]^k$ & $[1.1\pm0.3(-8)]^a, [3.0(-7)]^b$
\enddata
\tablecomments\\{\footnote 
  \noindent G31 observation:$^e$ \cite{gora21} (used ALMA with $\theta_b=0.98^{''}-1.19^{''}$ and $\rm N_{H_2}= 1.53 \times 10^{25}$ cm$^{-2}$), $^f$ \cite{rivi17} (used SMA with $\theta_b=0.75^{''}-3.5^{''}$ and $\rm N_{H_2}=1.2 \times 10^{26}$ cm$^{-2}$), $^n$ \cite{numm98} (used SEST and IRAM with $\theta_b=10^{''}$ and $\rm N_{H_2}=1.6 \times 10^{23}$ cm$^{-2}$)\\
 \noindent  Sgr B2 observation: $^x$ \cite{bell14} (used ALMA with $\theta_b = 1.2^{''}-1.9^{''}$ and $\rm N_{H_2}= 5.6 \times 10^{24}$ cm$^{-2}$), $^m$ \cite{bell13} (used IRAM with $\theta_b = 10^{''}-25^{''}$ and $\rm N_{H_2}= 5.6 \times 10^{24}$ cm$^{-2}$ \citep{bell14})\\
 \noindent  Orion KL observation: $^l$ \cite{brou15} (used ALMA with $1.4^{''} - 1.9^{''}$ and $\rm{N_{H_2} = 2.0\times10^{24}}$ cm$^{-2}$ \citep{favr11}), $^k$ \cite{terc15} (used ALMA with $1.29^{''}-2^{''}$ and $\rm N_{H_2}= 2.0 \times 10^{24}$ cm$^{-2}$ \citep{favr11})\\
 \noindent G10 observation: $^a$ This work (used ALMA with $\theta_b=1.39^{''} -2.44^{''}$ and $\rm N_{H_2}=1.35 \times 10^{25}$ cm$^{-2}$),$^b$ \cite{rolf11} (used SMA with $\theta_b=0.28^{''}-4.27^{''}$ and  with $\rm N_{H_2}= 5 \times 10^{24}$ cm$^{-2}$),$^{*d}$ \cite{dick01} (used SEST \& NRAO for possible detection of propenal and upper limit for vinyl alcohol with $\theta_b=45^{''}$ and $\rm N_{H_2}=10^{23}$ cm$^{-2}$)}
\end{deluxetable*}

\subsection{Binding Energy and Proton Affinity} \label{sec:BE_PA}
 A sizable portion ($\sim 60-70\%$ of the surface coverage) of the interstellar icy layers may contain water molecules. That is why the BE of the interstellar species is usually explored with the H$_2$O surface. However, the rest ($\sim30-40\%$) of the grain mantle would comprise other impurities.  CO$_2$, CO, and CH$_3$OH can occupy a sizable portion of the grain mantle in different lines of sight. \cite{kean01} summarized the relative abundance of CO, CO$_2$, and CH$_3$OH relative to water ice could vary in the range of $0.4-15$, $0.17-21$, and $1.5-30$, respectively, in different lines of sight. So, the BE with these surface species is also important in the various evolutionary phases. It is needed to construct a model which can consider the BE depending on the surface composition of the ice. 
\cite{furu18} reported a model in which the BE depends on the surface composition of
ice mantles.
However, showing the results of such a chemical model is out of scope for this work.

A realistic estimation of the BEs with various substrates is not available. In the absence of any experimental values, the quantum chemical method provides an educated estimate \citep{das18,wake17}.
Here, CO and CH$_3$OH molecules are considered a substrate for this purpose.

Gaussian 09 suite of programs employed for the quantum chemical calculations is utilized to evaluate the BE values. To calculate the BE, which is opposite to interaction energy for the bound system, the optimized energy for the complex system (where a species is placed at a suitable distance from the grain surface) is subtracted from the total optimized energies of the grain surface and species. To find the optimized energy of all structures, a Second-order M\o ller-Plesset (MP2) method with an aug-cc-pVDZ basis set \citep{dunn89} is used following \cite{das18}. The ZPVE and basis set superposition error (BSSE) correction are not considered for BE calculations.

Obtained BE values are shown in Table \ref{tab:substrate}.
It is interesting to note that a higher BE
for the alcohol is obtained in most cases than their corresponding aldehyde, ketone, or  ketene. Obtained values with CO and CH$_3$OH significantly differ from that of water. 
In terms of average values, the BE values with water monomer is $\sim 2.85$ times higher than the CO monomer and $1.16$ times lower in comparison to CH$_3$OH monomer.
Here, in our chemical model, the obtained BE values with CO and CH$_3$OH are not included because it is not available for the other species.
However, BE values obtained with the c-tetramer water configuration are included in our model with an appropriate scaling factor suggested in \cite{das18}.  In Table \ref{tab:substrate}, the ground state spin multiplicity of the species used to calculate the BE is also noted.  Separate calculations (job type ``opt+freq" in Gaussian 09 suite) with different spin multiplicities are used. The lowest energy electronic state in between these is noted as the ground state.

The protonation reactions (i.e., X + H$^+$ $\rightarrow$ XH$^+$) and proton transfer reactions (i.e., X + YH$^+$ $\rightarrow$ XH$^+$ + Y) have a significant impact on the 
interstellar chemistry \citep{wake10,herb86}. The ab-initio quantum chemical approaches provide a reliable value of the proton affinity for small molecules where the experimental determination of these parameters is a difficult task. A systematic study to calculate the  adiabatic PA of aldehydes, ketone, ketene, and corresponding alcohols is considered in this work.
PA is defined as the energy released when a proton is added to a system. Thus the energy difference between a species and a species with an additional proton (H$^+$) is estimated. It is calculated as the difference in energy (electronic + zero-point energy) between
a neutral species and its protonated analog, i.e.,
\begin{equation}
    PA = E_X-E_{XH^+},
\end{equation}
where $\rm{E_X}$ is the optimized energy of the species X, and $\rm{E_{XH^+}}$ is the optimized energy of the protonated species, XH$+$.
To find the optimized energy of all structures, a Second-order M\o ller-Plesset (MP2) method with an aug-cc-pVDZ basis set is again used. This is then verified via harmonic frequency calculations with the equilibrium geometries having only real
frequencies. Zero-point correction to energies is not considered to evaluate the PA of the species reported here.
The protonation of a neutral
species can occur in more than one position, which yields different protonated species with other PA. To avoid any ambiguity, only the highest PA is noted in Table \ref{tab:pa}. Protonated form with higher PA of a neutral species is found to be more stable and more abundant (in some cases), and thus may be detectable. In contrast, another protonated form with lower PA forming from the same neutral species is less stable. This species is very reactive, and the high reactivity of this protonated species may reduces its interstellar abundance. We notice that
ethenone has the highest and methanal has the lowest PA among all the species considered
for this study.

\begin{figure}
\centering
\includegraphics[height=9cm,width=8cm,angle=270]{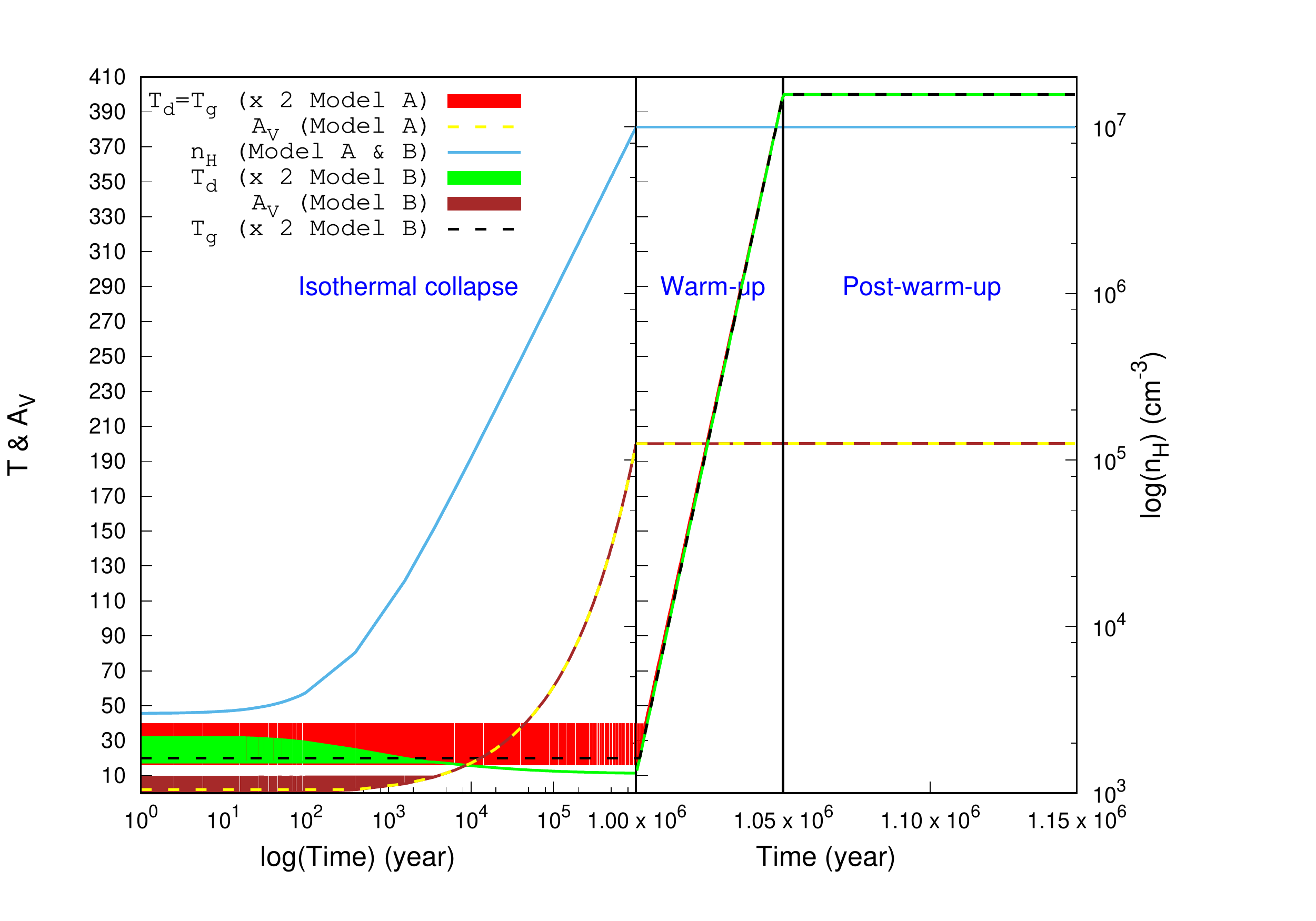}
\includegraphics[height=9cm,width=8cm,angle=270]{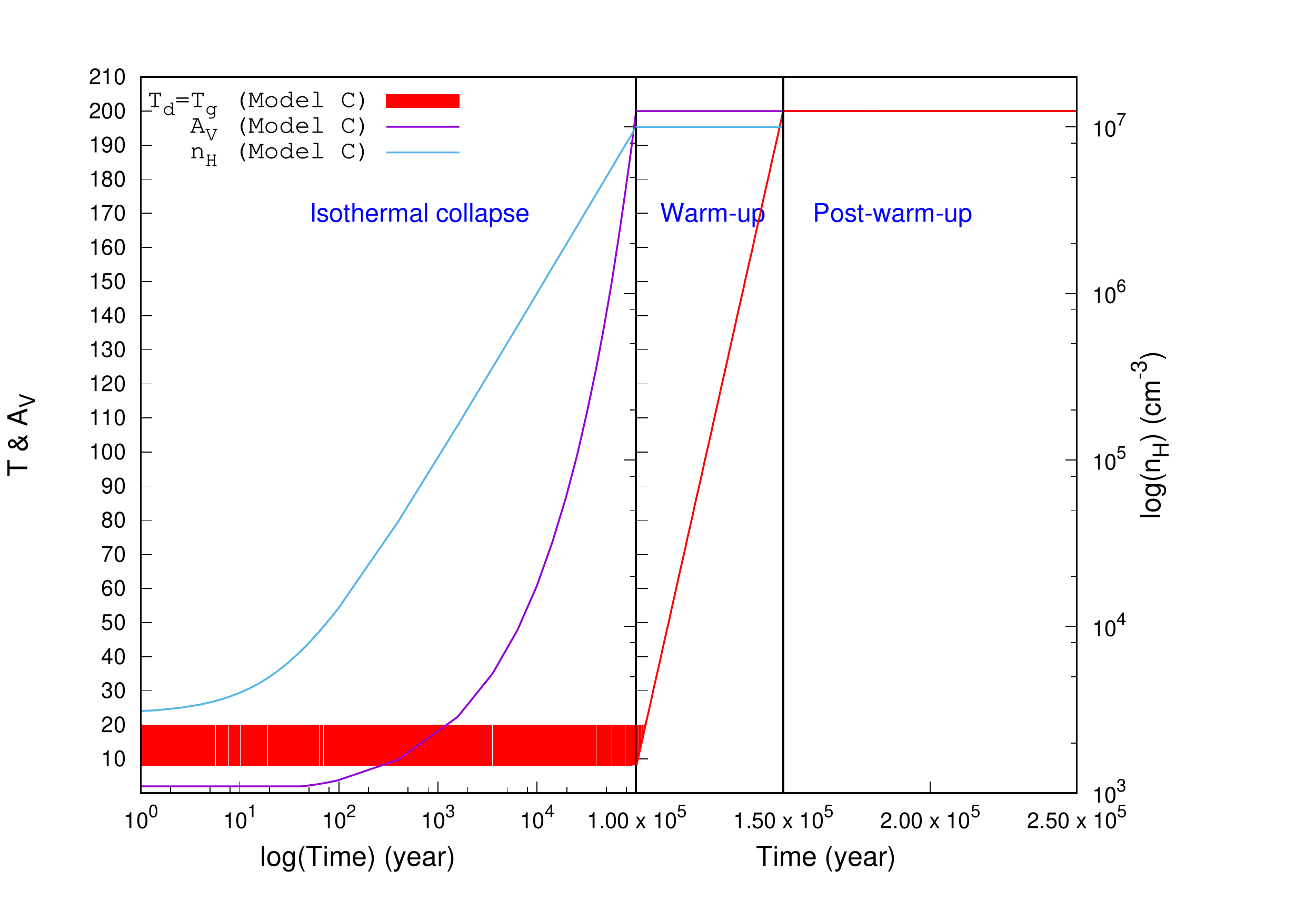}
\caption{Physical conditions considered in Model A and Model B are shown in the upper panel and  in the lower panel, we show the adopted physical conditions for Model C.
The X-axis of the figure shows the logarithmic scale for the isothermal collapsing phase, whereas, beyond that, it is shown on the regular scale.
In case of Model A and Model C, an isothermal collapsing
phase is considered whereas in Model B, a pre-stellar core phase is considered where temperature is decreased in the upper panel, temperature is scaled by a factor of 2 for the better visibility.
\label{fig:phys}}
\end{figure}
\section{Chemical modeling} \label{sec:chemical_model}
The Chemical Model for Molecular Cloud (hereafter CMMC) \citep{das15a,das15b,das19,das21,gora17a,gora17b,sil18,bhat21,gora20,sil21} is implemented for studying the formation of
aldehydes, ketone, ketene and their corresponding alcohols.
The low metallic elemental abundances are used as the initial abundances \citep[EA1 set]{wake08}.
Gas-phase pathways of CMMC are mainly adopted from the UMIST database \citep{mcel13}. Additionally, a complete network for the formation of alcohols and aldehydes is considered.
For the construction of these reaction sets, the ice phase reactions noted in Table \ref{tab:reac} are considered. The activation energy barriers calculated with the CCSD(T)/aug-cc-pVTZ level are adopted. Since, for the reaction 13a of Table \ref{tab:reac}, a true TS was not obtained with the CCSD(T)/aug-cc-pVTZ level of theory, the activation barrier obtained with the B3LYP/6-31+G(d,p) level of theory is used.
For the gas phase destruction of these species, interstellar photoreactions, cosmic-ray induced reactions, and ion-molecular reactions with  major ions (He$^+$, $\rm{H_3}^+$, HCO$^+$, H$_3$O$^+$, C$^+$, etc.) are considered. The destruction of some of these reactions was already available in \cite{mcel13}. If it is not available, a very similar reaction with the same rate constants is adopted for these reactions.
A cosmic ray rate of $1.3 \times 10^{-17}$ s$^{-1}$ is considered in all our models. 
Cosmic ray-induced desorption and chemical desorption with the efficiency
of $1\%$ is considered. For the all-grain surface species, a photodesorption rate of
$3 \times 10^{-3}$ per incident UV photon \citep{ober07} is adopted.
A sticking coefficient of $1.0$ is considered for the neutral species. Sticking coefficients of H and H$_2$ is
taken from \cite{chaa12}.
On the grain surface, the diffusive surface reaction can lead to chemical complexity.
However, to successfully continue the diffusive reaction mechanism, the reactants should be in the close
vicinity until they react. There is a possibility that two reactants can move apart or be desorbed back to the gas phase. So it is essential to consider the competition between diffusion, desorption, and reaction. Following \cite{garr11}, here this possibility is included. 
 At the high density ($\sim 10^7$ cm$^{-3}$) and around $10$ K, a notable portion of the grain surface would be covered by molecular hydrogen. At this stage, encounter desorption of H$_2$ \citep{hinc15,chan21,das21} needs to be considered to avoid an unnecessary surge in the grain’s molecular hydrogen. The grain mantle’s chemical composition at this phase is crucial to delivering the surface species into the gas phase during the warm-up stage.
Here, the encounter desorption mechanism of H$_2$ is considered in our model.

The BE ($E_d$) of the surface species plays a decisive role in controlling the chemical complexity of the interstellar ice.
A straightforward relation between the diffusion energy ($E_b$) of a species with the BE is considered by following
$E_b=R E_d$, where, R may vary in between $0.35-0.8$ \citep{garr07}. Here,  $R=0.35$ and $0.50$ are used to show their effects on the chemical complexity. These BEs are either taken from the KIDA database\footnote{\url{http://kida.astrophy.u-bordeaux.fr/}} or taken from \cite{das18}.
\cite{das18} provided BE values for some relevant interstellar species with the water surface. 
They computed the interaction energies, varying the size of the substrate. They showed that their method would yield a
minimum deviation with the experimentally obtained BE values when a higher-order cluster is used.
Table \ref{tab:substrate} shows the computed BEs of these aldehydes, alcohols, ketone, and ketene with various substrates. 
The ground state used to calculate the BE of these
molecules is also pointed out along with the BE values in Table \ref{tab:substrate}. 
\cite{das18} suggested a scaling factor $\sim 1.188$ while using the tetramer water configuration for the computation of the BE. The scaled BE values are also noted in Table \ref{tab:substrate}. 
Except for the glycolaldehyde and ethylene glycol pair, for all the cases, obtained BEs with the tetramer configuration are noted to be comparatively higher than that of their related aldehyde, ketone, or ketene. 
These new sets of BEs with the tetramer configuration of water molecules are included in our chemical model with appropriate scaling.

 Depending on the BEs and R values, four sets of BEs are constructed (see Table \ref{tab:setBE}). For set 1 and set 2, BEs are used from the KIDA database \citep{wake17}. The only difference between set 1 and set 2 is that for set 1, $R=0.5$ is used, whereas, for set 2, $R=0.35$ is used. 
Recently, \cite{das18} provided a set of BEs for $100$ relevant interstellar species with the
water surface. They used the tetramer configuration of water molecules for this calculation. They suggested
using a scaling factor of $1.188$ with this to estimate the BEs of these species. 
The set 3 is constructed with the BE values provided by \cite{das18}. 
\cite{sil17} reported the BEs of H and H$_2$ for a comprehensive collection of surfaces (benzene, silica, and water). 
They showed that the BE of H and H$_2$ are very sensitive to the choice of substrates. 
These variations should be expected for all the species if various substrates would have been considered.
But showing all the changes here would be out of  scope for this work. 
Here, one additional BE set is considered, where the BE of H and H$_2$ are kept the same as it was used in KIDA, but for the others, it is the same as it is in set 3. Thus, in set 3, BE of H, and H$_2$ of 148 K, and 627 K, respectively, are used from \cite{das18} after scaling by 1.188, whereas in set 4, it is used 650 K and 440 K, respectively for H, and H$_2$ \citep{wake17}.
$R=0.35$ is considered for both the set 3 and 4. 

\begin{table}
\scriptsize
\centering
    \caption{Different sets of BE used based on the ratio between the diffusion energy to binding energy (R).}
     \begin{tabular}{ccc}
    \hline
         &BEs used& R\\
      \hline
      set 1&\cite{wake17}&0.5\\
      set 2 &\cite{wake17}&0.35\\
      set 3 & 1.188 $\times$ BE of \cite{das18}&0.35\\
      set 4 &1.188 $\times$ BE of \cite{das18}, BE of&0.35\\
      & H, and H$_2$ from \cite{wake17}  &  \\
      \hline
    \end{tabular}
    \label{tab:setBE}
\end{table}

\begin{figure*}
\centering
\includegraphics[height=18cm,width=14cm,angle=270]{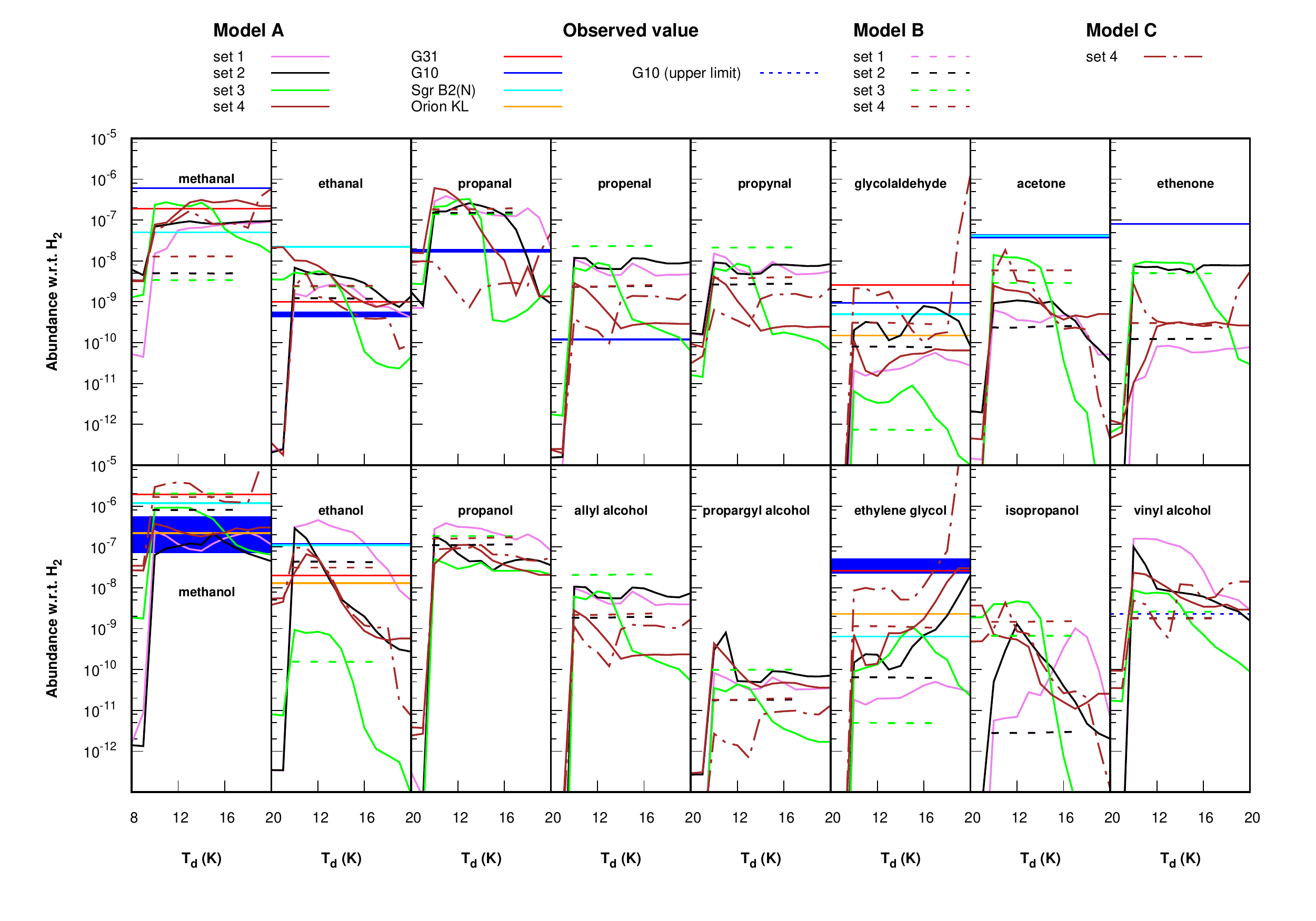}
\caption{The initial dust temperature versus the peak gas-phase abundance obtained from Model A, Model B, and Model C is shown in various cases. 
Eight pairs of species are shown. Observed/upper-limit of the abundances of these species are marked with red (G31), blue (G10),  Sgr B2(N) (cyan), and Orion KL (orange) if available. Since, in G10, some species are identified here, observed errors are shown with the thick blue curve. The solid
curve represents the cases with Model A and dashed curves represent the same cases for Model B. Results obtained with Model C and set 4 are shown with the dashed-dot lines. 
The lower eight panels show
the peak abundance variation of alcohols whereas the top panels show their related aldehyde, ketone, and ketene respectively.
Thus, we conclude that there is both observational and theoretical evidence for production of alcohols via hydrogenation of precursor aldehydes, ketones and ketenes.
\label{fig:evolution}
}
\end{figure*}

\begin{figure*}
\centering
\includegraphics[height=16.5cm,width=9cm,angle=270]{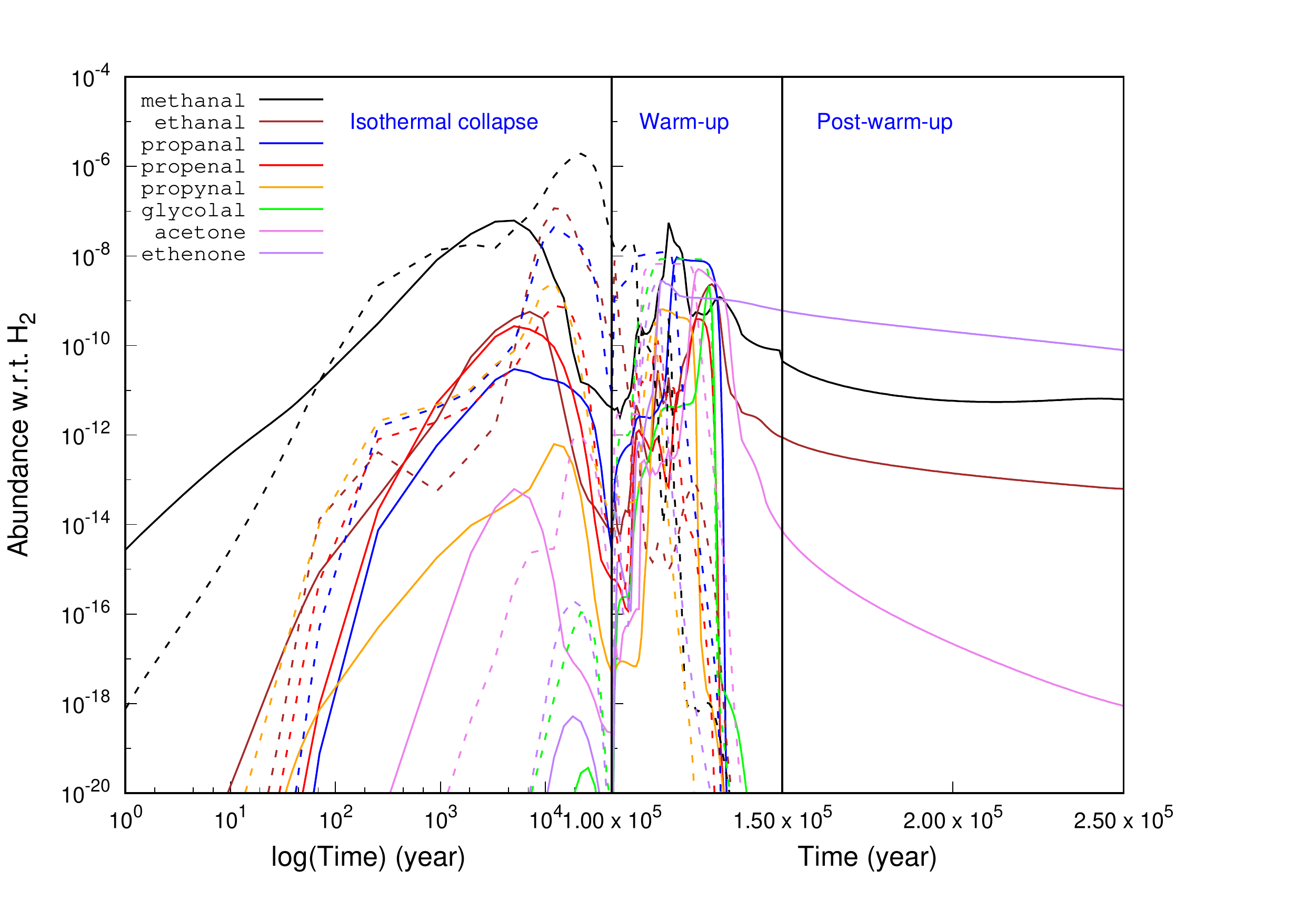}
\includegraphics[height=16.5cm,width=9cm,angle=270]{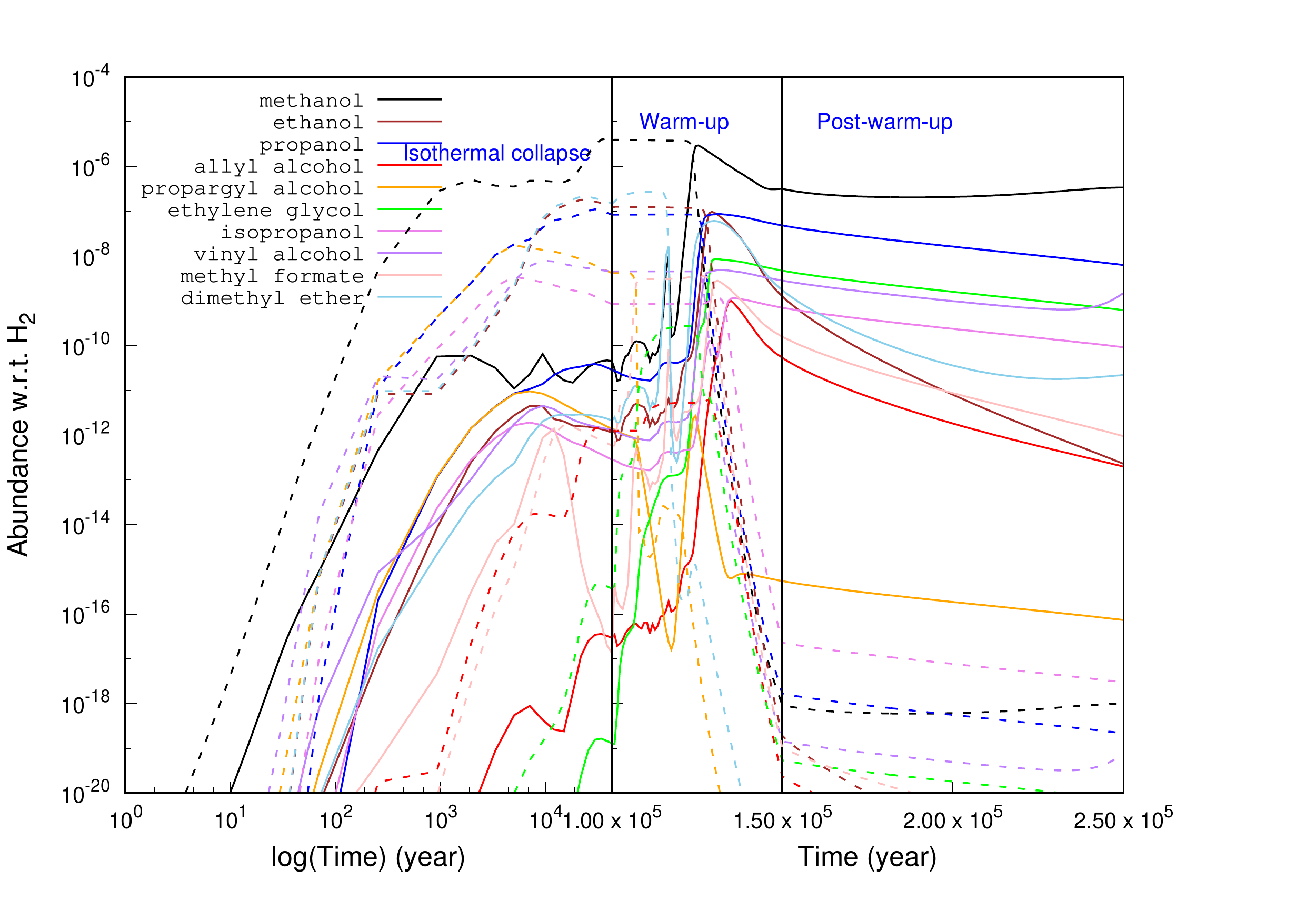}
\caption{The chemical evolution of aldehyde, ketone, ketene and their corresponding alcohols are shown with the
set 4 of CMMC Model C while initial dust temperature kept at $10$ K. 
The X-axis of the figure shows the logarithmic scale for the isothermal collapsing phase, whereas, beyond that, it is shown on the regular scale.
Gas phase abundance are represented
with the solid curve whereas for the ice phase it is with the dashed curve. Additionally, the time evolution of
propynal, methyl formate, and dimetyhyl ether are shown.
\label{fig:time-evo}}
\end{figure*}

\begin{figure*}
\centering
\includegraphics[height=18cm,width=12cm,angle=270]{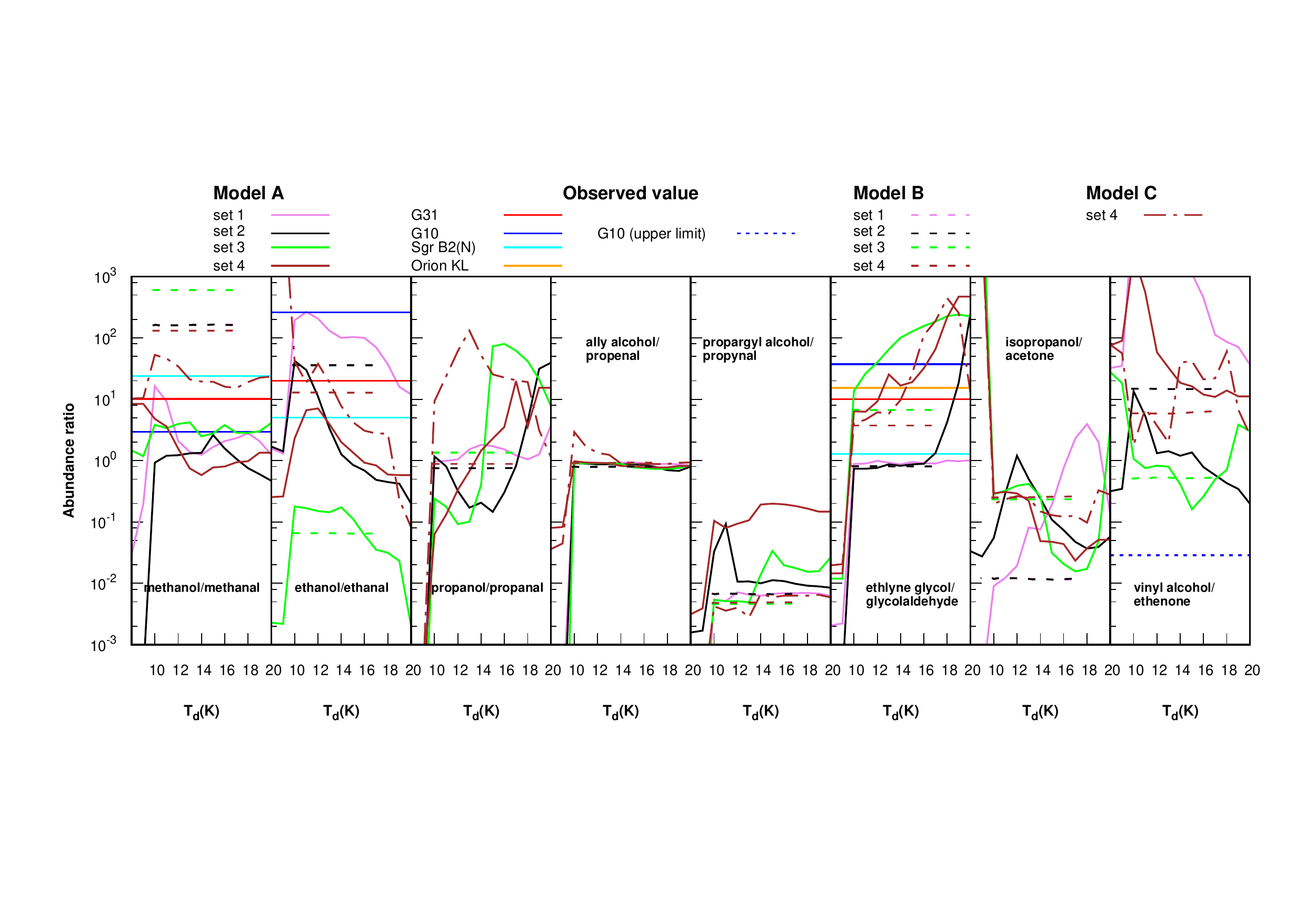}
\caption {The obtained ratio in between six pairs of alcohol and aldehyde along with the ratio 
of alcohol-ketone and alcohol-ketene is shown. The observed ratio is marked in red (G31), blue (G10), cyan (Sgr B2(N)), and orange (Orion KL). Model A results are highlighted with  solid curves whereas the same with Model B is represented with dashed curves and with Model C is represented with dot-dashed curves.
\label{fig:ratio}}
\end{figure*}

\subsection{Physical conditions}
Here, a physical condition is used, which is suitable to explain the properties of a hot core.
Three models (Model A, Model B, and Model C ) are implemented to describe the chemical complexity around this region.
The physical condition of our Model A is divided into three distinct phases.
In the first phase (10$^5$-10$^6$ years),
an isothermal (gas and grain temperature are kept the same) collapse of the cloud is considered from
a minimum hydrogen number density (${n_H}_{min}$) $3 \times 10^3$ cm$^{-3}$ to a maximum hydrogen number
density (${n_H}_{max}$) 10$^7$ cm$^{-3}$. 
 During this phase, the visual extinction parameter ($A_V$) can gradually increase due to increased density.
The following relation between $A_V$ and n$_H$ is used to consider this situation \citep{lee96,gora20,das11}.
\begin{equation} \label{eqn:av}
A_V=max({A_V}_{min},(\sqrt{n_H/{n_H}_{min}}-1)/(\sqrt{{n_H}_{max}/{n_H}_{min}}-1) \times {A_V}_{max}).
\end{equation}
It is considered that $A_V$ can vary  from a minimum value (${A_V}_{min}=2$) to
a maximum value (${A_V}_{max}=200$.
In the second phase ($5 \times 10^4$ years), the cloud remains constant
at ${n_H}_{max}$, ${A_V}_{max}$. At the end of this phase,
gas and dust temperature can reach up to $200$ K. The third or last phase is a post warm-up phase
(period of  $1.0 \times 10^5$ years),
where $n_H$, A$_V$, and temperature are kept fixed at their respective highest values. 

In the case of Model B, the pre-stellar collapsing phase is considered instead of the initial
isothermal collapsing phase. In this phase, the dust temperature is allowed to decrease as the collapse proceeds.
Gas temperature is kept constant at $10$ K during this collapse.
The other two phases are considered the same as Model A.
Dust temperature of the pre-stellar collapsing phase is calculated by using the
following empirical relation proposed by \cite{hocu17} and used in \cite{shim20}:
\begin{equation}
{T_d}^{Hoc} = [11 + 5.7 tanh(0.61 - log_{10} (A_V))] {\chi}_{uv}^{1/5.9},  
\label{eqn:hock}
\end{equation}
where $\chi_{uv}$ is the Draine UV field strength \citep{drai78}, corresponding to
$1.7G_0$  using the Habing field \citep{habi68}.  Here,  G$_0$ is the Far Ultraviolet radiation field in the Habing unit ($\sim 1.6 \times 10^{-3}$ erg cm$^{-2}$ s$^{-1}$) integrated over the energy range $6 - 13.6$eV. Here, $\chi_{uv}=1$ is used.

Model C is the same as Model A. The only difference between Model A and Model C is the isothermal collapsing time scale of the first phase. In Model C, a shorter time scale ($10^5$ years) for the isothermal collapse is considered. It is more relevant for high mass star-forming regions. Thus, in the case of Model A, we have a density slope of $\sim 10$ cm$^{-3}$ per year, whereas it is $\sim 100$ cm$^{-3}$ per year in the collapsing phase. The  warm-up and  post-warm-up timescales are kept the same as in Model A. 

The physical condition described here are summarized in Figure \ref{fig:phys}. 
The time evolution of density, temperature, and visual extinction 
in the three distinct phases are shown for Model A, Model B, and Model C. 
The left axis of Figure \ref{fig:phys} represents the variation of temperature
and visual extinction parameter in the linear scale, whereas the right axis represents the density curve in the logarithmic scale. For the better visibility, in the top panel, temperature is plotted by scaling by a factor of 2.
Model A and Model B start with an isothermal collapsing phase. In this case, the dust temperature and gas temperature
are kept the same. Here, a region having an initial temperature between $8-20$ K is explored to check the sensitivity of our Model A
on the initial choice of the initial temperature.
The temperature range for the initial phase of Model A and Model C is highlighted with the red shaded region. The visual extinction for the initial phase (isothermal collapsing phase) of Model A and C
is kept constant at 2. In the case of Model B, the gas temperature of the 1$^{st}$ phase is kept fixed at $10$ K, and dust temperature is estimated
based on equation \ref{eqn:hock}. The initial visual extinction of Model B is varied in between $0.1-10$ 
(corresponds to an initial dust temperature of $16-5$ K from equation \ref{eqn:hock})
to check its effect on chemical evolution. 
The initial choice of visual extinction parameter and its estimated dust temperature of Model B are highlighted in green and yellow, respectively, in Figure \ref{fig:phys}. In the warm-up phase of Model A, Model B, and Model C, $A_V$ and n$_H$ are kept constant at their highest value, and the temperature
is allowed to increase up to $200$ K. It is the
typical temperature of a hot core \citep{garr08}. In the post-warm-up phase, both the models have identical conditions.

\subsection{Modeling results}
\subsubsection{Abundances}
In Model A, various cases starting with the initial temperature between $8-20$ K are considered.
Figure \ref{fig:evolution} shows the variation of peak abundances with respect to H$_2$ obtained during the simulation time scale. Six pairs of aldehyde-alcohol, one
pair of ketone-alcohol, and one pair of ketene-alcohol shown. 
Observed abundances in G10, G31, Sgr B2(N), and Orion KL are shown in solid blue, red, cyan, and orange curves, respectively.
In Figure \ref{fig:evolution}, the violet solid line represents the case with set 1 BEs reported in Table \ref{tab:setBE}. 
The solid black
line represents the case with set 2. 
A lower value of $R$ 
provides more mobility to the surface species keeping the same resident time on grains. 
It yields a dramatic change in the production of surface species. The changes in the chemical composition of
interstellar ice would reflect a significant difference between the gas phase abundances of the species shown
in Figure \ref{fig:evolution}. 
The case with set 3 is represented with the solid green line in Figure \ref{fig:evolution}. The set 4 case is with the solid brown curve in Figure \ref{fig:evolution}.
It is evident from Figure \ref{fig:evolution} that the resulting peak abundance is highly sensitive on the choice of initial temperature. More
specifically, on the dust temperature. Some cases of Model A are tested with a different gas temperature than dust and did not
find significant differences in the peak abundance. 

It is noticed that set 1 and set 2 BEs are helpful to explain the abundances of these aldehydes and alcohols except glycolaldehyde and ethylene glycol. The set 3 BEs deviate very much from the
observed value. It is mainly because of the adopted BE values of H (148 K) and H$_2$ (627 K). 
However, with the set 4 BEs, the observed abundance of ethylene glycol could be explained. It is because set 4 considered the BE of H and H$_2$  same as these are in set 1 and set 2 ($650$ K and $440$ K, respectively for H and H$_2$). 
The ice phase origin of these species is evident from the results obtained from Model A.
The results obtained from Model A seem to be highly sensitive to the initial dust temperature, which suggests that either these species were formed during the initial cold phase or the seeds of these species were produced during the initial stage.
The final gas-phase abundance of these species is highly affected by the surface species present at the
beginning of the warm-up phase. The surface species at the warm-up phase are re-processed with the elevated
temperature, where radicals take an active part in building the chemical complexity. Once the warm-up phase
starts, the desorption of the surface species enhances and might reach their respective sublimation temperatures and populate
the gas phase. With set 4 BE and Model A, it is possible to explain the observed abundances except glycolaldehyde. 

The results obtained with Model B are also shown in Figure \ref{fig:evolution}. All the four cases mentioned in the context of Model A are also considered with Model B. To visualize the differences between 
Model A and Model B, Model B cases are highlighted with the dashed curves, but with the same color codes.
It is interesting to note that for Model B, significant changes of abundance for the adopted BE parameters are noticed. Like Model A, the results of Model B are not so much sensitive to the initial choice of dust temperature. 
The reason behind this deviation is the adopted dust temperature at the 1$^{st}$ phase of Model A and Model B (see Figure \ref{fig:phys}). In Model A, the dust remains at a
constant temperature for $\sim 10^6$ years, whereas in Model B, in between a few times $10^3$ years, it drops down to
temperature $<10$ K. This affects the mobility of the surface species. Since it stays at $<10$ K during its first phase of evolution, Model B does not show dramatic changes between the variation of the initial dust temperature. 

The orange curve of Figure \ref{fig:evolution} represents the variation of the peak abundances of these species with set 4 and Model C. It is interesting to note that with the shorter collapsing time scale (relevant for the hot core) of Model C, peak abundances of complex organic molecules are significantly enhanced. Most of the observed abundances could be explained with set 4 of Model C.
The increase in the abundances of the complex organic molecule with the shorter collapsing time is not so straightforward. Many parameters are involved in it.  But the shorter collapsing time leads to a much steeper slope for the density, which means that the depletion will be much faster and the complex molecules, once produced, will have a comparatively shorter time to be destroyed further.
In Figure \ref{fig:time-evo}, the time evolution of the gas phase and ice phase abundances of aldehydes, alcohols, ketone, and ketene with respect to H$_2$ are shown. Only the results obtained with Model C by considering an initial gas and ice temperature of $10$ K is shown.
There is some uncertainty in the observations. Additionally, 
the chemical model includes lots of estimated parameters (BEs, reaction pathways, reaction rates, interaction with dust, etc.) that can induce severe skepticism. 
Due to these reasons, it is not easy to match all the observed abundances simultaneously.
In Figure \ref{fig:evolution}, the peak abundance is taken 
beyond the isothermal collapsing phase.
Figure \ref{fig:time-evo},
depicts that beyond the isothermal phase these peak gas-phase abundance appeared during the warm-up phase, when the temperature varies in between $\sim 100-120$ K. So our time uncertainty is relatively small ($\sim 10^4$ years).

A comparison between our obtained gas-phase peak abundance with the observational results is shown in Table
\ref{tab:comp}. Only the results obtained with set 4 of Model C are used to show this comparison.
Minimum and maximum peak abundances obtained within the initial temperature range considered for Model C and set 4 are noted in Table \ref{tab:comp}.
The errors in obtained abundances are only shown for those species which are identified herein G10. The errors are calculated by taking the uncertainty in estimating the H$_2$ column density and the column density obtained from rotation diagram analysis. \cite{gora20} obtained an average column density of $\sim 1.35 \times 10^{25}$ cm$^{-2}$ . The error in estimating the column density was $\pm 1.0 \times 10^{22}$ cm$^{-2}$, which mainly arose from the uncertainty of the measured flux.  The error in estimating the column density of the species is taken from Table \ref{tab:rottmp}. For estimating the abundance errors for glycolaldehyde, only the uncertainty in predicting the H$_2$ column density is considered.

It is noticed that our results
can successfully explain the observed abundances.

\subsubsection{Molecular ratios}
In Figure \ref{fig:ratio}, the temperature variation of the ratio obtained between the peak abundance of six pairs of alcohol-aldehyde, one pair of alcohol-ketone, and one pair of
alcohol-ketene for Model A (solid curve), Model B (dashed curve), and Model C (dot-dashed curves) are shown. 
The observed ratios for G31, G10, Sgr B2(N), and Orion KL are shown with red, blue, cyan, and orange solid lines, respectively. The simulated ratio between the peak abundance of alcohol and aldehyde/ketone/ketene obtained with set 4 and Model C, along with the observed abundances of these species is noted in Table \ref{tab:comp}. 
 Table \ref{tab:comp} depicts the observed abundances in G31, G10, Sgr B2, and Orion KL, along with our modeled abundance.  Mainly the interferometric observations (wherever available) are noted to facilitate a comparative study between the chemical complexity of these hot cores. However, it is worth pointing out that all these observations are carried out with different facilities with very diverse angular and spatial resolutions. The facility used for each observations along with the H$_2$ column density used to derive the abundances are noted at the footnote of Table \ref{tab:comp}. 
Here, the band 4 archival data of G10 is only analyzed, which reports only one alcohol-aldehyde pair (ethylene glycol-glycolaldehyde pair). These two species are recognized for the first time in this source. Earlier, \cite{rolf11} obtained the methanol-methanal pair. 
Any transitions of methanal are not identified within our spectral bands. So, a direct comparison between our obtained ratios and others could not be performed. For example, Table \ref{tab:comp} denotes a higher methanol abundance obtained by \cite{rolf11} than ours. This is because the beam size of our band 4 observation of ALMA varies between $1.39^{''}-2.44^{''}$ ($\sim 12000 - 21000$ AU, considering 8.6 kpc distance). In contrast, the beam size of the submillimeter Array (SMA) observation at 345 GHz of \cite{rolf11}  is 0.3$^{''}$ ($\sim 2600$ AU, considering 8.6 kpc distance). The integrated emission map of the methanol transitions shown in Figure \ref{fig:maps3} suggests that the methanol is more compact.  So as expected, with the higher spatial resolution, \cite{rolf11} can trace the more inner region of the G10, and thus they obtained a higher abundance than us. 
Furthermore, for deriving the abundances from \cite{rolf11}, a hydrogen column density of $5 \times 10^{24}$ cm$^{-2}$ is used. In contrast, our abundance is derived by considering $1.35 \times 10^{25}$ cm$^{-2}$ \citep{gora20}, which might also create significant variation. 
Since our chemical model does not consider the spatial variation of the abundances, explaining the observed ratio with the suitable beam size is also out of scope for this work.

The observed ratio is calculated from the observed pair from the same source (if available).
It depicts that the methanol to methanal abundance ratio in these sources can vary between $2.95-23.8$, whereas our calculated methanol to methanal ratio varies between $10-53$. The observed ratio between ethanol to ethanal varies between $5-261$, whereas our model results show it ranges between $0.08-25600$. The observed ethylene glycol to glycolaldehyde ratio varies between $1.3-37$, whereas our model derives $0.019-445$. The observed ratio between the vinyl alcohol to ethenone is $0.021$, whereas our chemical model estimates $1.78-77$. In brief, the observed ratio of three alcohol-aldehyde pairs (methanal-methanol, ethanal-ethanol, glycolaldehyde-ethylene glycol) is  $>$ 1 in G10, G31, Sgr B2(N), and Orion KL.  
Thus, in the high-mass star-forming regions, the abundances of these alcohols are generally higher than their respective aldehydes.
Our estimated molecular ratios for these alcohol-aldehyde pairs are similar to the observed values in the temperature range $12-18$ K (see Figure \ref{fig:ratio}).
However, in the case of the vinyl alcohol to ethenone, the observed abundance ratio could not be explained. It could be because of the consideration of the upper limit of vinyl alcohol observation from the single-dish observation of \cite{dick01} for deriving the molecular ratio. 
 A little ambiguity between our obtained ratio of the dimethyl ether (DE) and methyl formate (MF) is obtained. DE/MF $<$1 is received from our observational analysis, whereas \cite{rolf11} noted it $>$1 for the same source. Also, in other high mass sources, the ratio is $>$1 (4.5 for Sgr B2, 2 for G31, and 2.5 for the Orion KL).  Our modeling results find that the ratio of DE with the MF can vary between $0.12-152$.

\section{Conclusions} \label{sec:conclusion}
Here, our CMMC model coupled with the observational study is used to explain the abundance of some interstellar aldehydes, ketone, ketene, along with their corresponding alcohols. 
It is noticed that there is both observational and theoretical evidence for production of alcohols via hydrogenation of precursor aldehydes, ketones and ketenes.
The following are the
main highlights of this paper.

\begin{itemize}
\item 
  Various aldehydes, alcohols, and a ketone are identified in G10. Among them, ethylene glycol and propanal are detected for the first time in G10, and glycolaldehyde is tentatively detected.

\item
The column densities and fractional abundances of the observed molecules are estimated assuming LTE conditions. The kinetic temperatures of the gas are also evaluated, which varies from 72 K to 234 K. 

\item The spatial distributions of the observed species are investigated, and their zeroth-order moment maps and emitting regions are provided. However, molecular emissions are not  well spatially resolved or, at best, marginally resolved. Thus, we need high angular resolution data to understand their detail distributions in G10. 

\item Successive hydrogenation reaction can lead to the formation of alcohol starting
from the aldehyde/ketone/ketene. 
Extensive quantum chemical calculations are carried out to yield the TS of some of these reactions.
The first step of the hydrogen addition reactions has an activation barrier, whereas the second step could be considered barrier-less radical-radical reactions.  Our higher-level (CCSD(T)/aug-cc-pVTZ) ice-phase calculations depict the first step of hydrogenation reaction at the carbon atom position for all the pairs (except acetone-isopropanol) is comparatively favorable than the oxygen atom position. In the acetone-isopropanol couple, an actual TS could not be obtained for the hydrogenation to the carbon atom position. However, the lower-level calculation (DFT-B3LYP/6-31+G(d,p)) in the ice-phase shows the same trend.

\item BEs of these species are computed with various substrates such as H$_2$O, CH$_3$OH, and CO. 
For most cases, the BE of alcohols is found to be greater than that of the corresponding aldehyde/ketone/ketene.
On average, the computed BEs with CH$_3$OH monomer is found $\sim 1.16$ times higher, and with CO monomer, it is $\sim 2.85$ times lower than that of the water monomer. These new sets of BEs would help understand the chemical complexity, where CO and CH$_3$OH would also efficiently cover the icy mantle. BEs of the species reported in this study for the alcohol, aldehyde, ketone, and ketenes is a little bit on the higher side ($\sim  3000$ K) with the H$_2$O c-tetramer configuration. Comparatively, a higher surface temperature is needed to transfer these surface species in the gas phase. Such a high temperature is usually achieved deep inside the cloud in the hot core phase. If the grain surface pathways are the only deriving route, then species with lower BE are expected to have more extended emission than the species with relatively higher BE. Since our observed beam size was large, it is not possible to spatially resolve the inner region and comment on the dependency between the BE and size of the emitting region.

\item Our CMMC model with the shorter collapsing time scale ($\sim 10^5$ years) was able to reproduce the observed abundance and observed ratio between various species.

\end{itemize}

\acknowledgments
This paper makes use of the following ALMA data: ADS/JAO.ALMA$\#$2016.1.00929.S. ALMA is a partnership of ESO (representing its member states), NSF (USA) and NINS (Japan), together with NRC (Canada), MOST and ASIAA (Taiwan), and KASI (Republic of Korea), in cooperation with the Republic of Chile. The Joint ALMA Observatory is operated by ESO, AUI/NRAO and NAOJ.
 S.K.M.  acknowledges CSIR fellowship (Ref no. 18/06/2017(i) EU-V). P.G. acknowledges support from a Chalmers Cosmic Origins postdoctoral fellowship. M.S. gratefully acknowledges DST, the 
Government of India for providing financial assistance through the DST-INSPIRE Fellowship [IF160109] scheme. 
A.D. acknowledges ISRO respond project (Grant No. ISRO/RES/2/402/16-17) for partial financial support. J.C.T. acknowledges support from the Chalmers Foundation and VR grant. T.S. and A.D. acknowledge Indo-Japan Collaborative Science Programme.
This research was possible in part due to a Grant-In-Aid from the Higher Education Department of the Government of West Bengal.

\software{Gaussian 09 \citep{fris13}, CASA 4.7.2 \citep{mcmu07}, CASSIS \citep{vast15}.}

\clearpage

\appendix
\restartappendixnumbering
\section{emitting diameter}
The emitting diameter of each transition of the observed molecules is shown in Table \ref{tab:dia}. Here, the average emitting diameter refers to the average source size.  2D Gaussian fittings of a region (area enclosing $50\%$ line peak) of the moment map image are performed.  After averaging the semi-minor and semi-major axis, the emission region's diameters are obtained.
\startlongtable
\begin{deluxetable*}{ccccc}
\tablecaption{The emitting diameter of each transition of observed species toward G10.47+0.03. \label{tab:dia}}
\tablewidth{0pt}
\tabletypesize{\footnotesize}
\tablehead{
\colhead{Species} & \colhead{${\rm J^{'}_{K_a^{'}K_c^{'}}}$-${\rm J^{''}_{K_a^{''}K_c^{''}}}$} & \colhead{Frequency} & \colhead{E$_u$} &\colhead{Emitting diameter (average)} \\
\colhead{ } & \colhead{ } & \colhead{(GHz)} & \colhead{(K)} & \colhead{$\theta_s^{''}$ }}
\startdata
Methanol 
&$\rm{11_{0,11}-11_{1,11}}$ E, vt=0& 154.425832$^*$&166.1&1.81\\
($\rm{CH_3OH}$)&$\rm{12_{0,12}-15_{1,12}}$E, vt=0& 153.281282$^*$&193.8&1.79\\
&$\rm{15_{0,15}-15_{1,15}}$E, vt=0&148.111993&290.7&1.78\\
&$\rm{7_{-1,7}-6_{0,6}}$E, vt=1&147.943673&356.3&1.56\\
&$\rm{17_{-4,13}-18_{-3,16}}$E, vt=0&131.134094&450.9&1.45\\
&$\rm{21_{0,21}-21_{1,21}}$E, vt=0&129.720384&546.2&1.33\\
&8$\rm{8_{-7,1}-7_{-6,1}}$E, vt=1&153.128697&664.5&1.23\\
&&&&\\
Ethylene Glycol&$\rm{8_{4,5}-7_{3,5}}$, 1-0&153.567383&25.4&1.83\\
(g-HOCH$_2$CHO)&$\rm{13_{3,11}-12_{2,10}}$, 0-0&159.768239&48.8&1.85\\
&$\rm{13_{9,4}-12_{9,3}}$, 0-1&130.998583&83.8&1.04\\
&$\rm{27_{3,25}-27_{2,26}}$, 0-0&148.082465&185.5&1.04\\
&$\rm{28_{7,21}-28_{6,23}}$, 1-0&131.229156&223.4&1.21\\
&$\rm{33_{10,23}-32_{11,22}}$, 1-1&130.115657&323.0&1.60\\
&$\rm{39_{8,32}-39_{7,33}}$, 1-1&153.325448&414&1.27\\
&&&&\\
Acetaldehyde&$\rm{8_{3,6}-7_{3,5}}$,A&154.2746864&53.7&1.49\\
($\rm{CH_3CHO}$)&$\rm{8_{3,5}-7_{3,4}}$,E&154.296489&53.7&1.53\\
&$\rm{8_{4,4}-7_{4,3}}$, A&154.201471&69.5&1.38\\
&$\rm{8_{5,4}-7_{5,3}}$, A&154.161467&89.8&1.33\\
&$\rm{8_{3,5}-7_{3,4}}$, A&154.173895&114.4&1.0\\
&&&&\\
Propanal &$\rm{9_{7,2}-9_{6,3}}$, A&148.005042&49.6&1.18\\
($\rm{CH_3CH_2CHO}$)&$\rm{10_{7,3}-10_{6,4}}$, E&147.867178&54.6&1.27\\
&$\rm{16_{2,15}-15_{1,14}}$, E&159.932346&68.6&0.83\\
&$\rm{25_{3,22}-25_{2,23}}$, A&153.554511&172.7&1.87\\
&$\rm{27_{5,23}-27_{4,24}}$, E&154.066288&206.2&1.08\\
&&&&\\
Glycolaldehyde&$\rm{15_{0,15}-14_{1 14}}$&153.597995&60.5&1.26\\
($\rm{HOCH_2CHO}$)&$\rm{20_{7,13}-20_{6,14}}$&153.614231&147.0&0.89\\
&&&&\\
Acetone&$\rm{12_{2,11}-11_{1,10}}$, AA&130.924799&44.1&1.28\\
($\rm{CH_3COCH_3}$)&$\rm{11_{5,7}-10_{4,6}}$, EE&147.684364&48.4&1.83\\
&$\rm{13_{2,11}-12_{3,10}}$, AE&149.395864&55.8&1.56\\
&$\rm{14_{3,12}-13_{2,11}}$, EE&159.2476343&63.4&1.29\\
&$\rm{22_{4,18}-22_{3,19}}$, EE&160.118153&157.1&1.38\\
&$\rm{24_{6,18}-24_{5,19}}$, EE&159.415955&198.9&1.50\\
&$\rm{25_{10,15}-25_{9,16}}$, AE&130.708968&241.0&1.0\\
&$\rm{26_{10,17}-26_{9,18}}$, EE&149.190766&251.0&1.35\\
&&&&\\
 Methyl formate &$\rm{11_{2,10}-10_{2,9}}$, A&130.016790&40.7&1.87\\
($\rm{CH_3OCHO}$)&$\rm{11_{2,10}-10_{2,9}}$, E&130.010105&40.7&1.80\\
&$\rm{12_{4,9}-11_{4,8}}$, A&148.805941&56.8&1.95\\
&$\rm{13_{3,11}-12_{3,10}}$, A&158.704392&59.6&1.85\\
&$\rm{13_{7,6}-12_{7,5}}$, A&160.178942&86.2&1.82\\
&$\rm{13_{7,6}-12_{7,5}}$, A&160.193496	&86.2&1.66\\
&$\rm{13_{10,4}-12_{10,3}}$, A&159.662793&120.0&1.67\\
&$\rm{12_{1,12}-11_{1,11}}$, A&131.377495&230.0&1.0\\
&$\rm{12_{4,8}-11_{4,7}}$, E&148.575217&244.1&1.69\\
&&&&\\
Dimethyl ether&$\rm{6_{1,6}-5_{0,5}}$, AA&131.406552&19.9&1.91\\
($\rm{CH_3OCH_3}$)&$\rm{9_{0,3}-8_{1,8}}$, AE&153.055196&40.4&1.83\\
&$\rm{11_{1,10}-10_{2,9}}$, AE&154.456506&62.9&1.95\\
&$\rm{24_{4,20}-24_{3,21}}$, EE&153.385902&297.5&1.44\\
\enddata
\tablecomments{ $^*$ optically thick}
\end{deluxetable*}

\clearpage
\restartappendixnumbering
\section{observed spectra toward G10}
\begin{figure*}
\centering
\includegraphics[width=18cm]{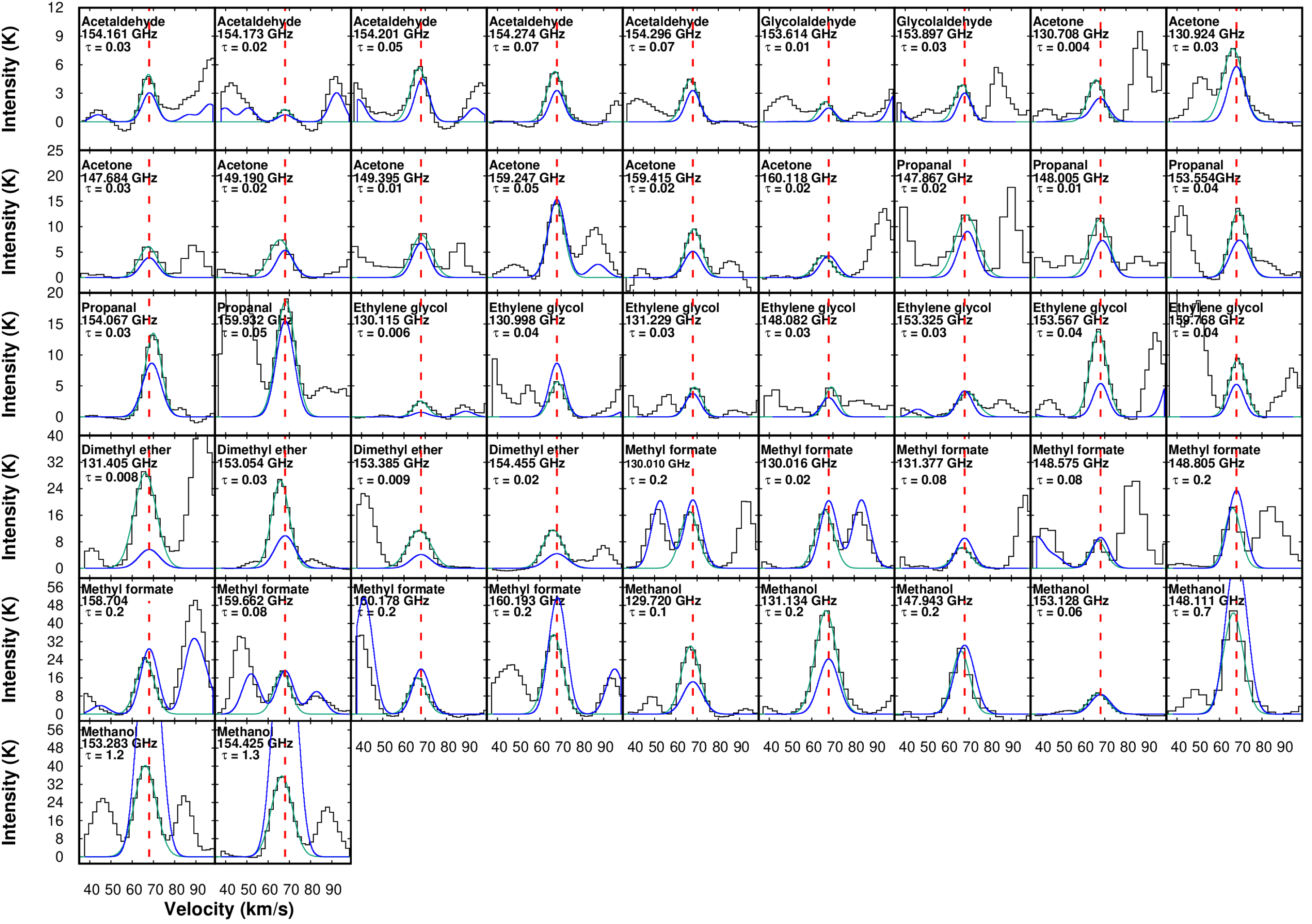}
\caption{ The observed spectra (solid black lines) along with the fitted Gaussian profile (solid green lines) and the LTE synthetic spectrum with our derived parameters (rotation temperature and column density) from rotational diagram analysis with the solid blue lines are shown. For the LTE calculation, a source size of 2$^{''}$ and average FWHM obtained from the Gaussian fitting (noted in Table \ref{tab:dataobs}) of each species are used. The dashed red line shows the systematic velocity (V$_{LSR}$) of the source at $\sim$ 68 km s$^{-1}$. In addition, the name of the species and their respective transitions (in GHz) and  optical depth of each transitions are given in each panel. Optical depth of all transitions are calculated based on the parameters assumed during LTE modeling. The last two transitions of methanol are optically thick, which are largely overproduced with the LTE consideration.
The intensity of all the transitions of dimethyl ether is underproduced, possibly because our overestimated FWHM for dimethyl ether. 
}
\label{fig:fit1}
\end{figure*}

 The observed, fitted and LTE spectra of observed transitions are shown in Figure \ref{fig:fit1}.

\clearpage
\restartappendixnumbering
\section{Line maps}
\begin{figure*}
\centering
\includegraphics[width=18cm,height=20cm]{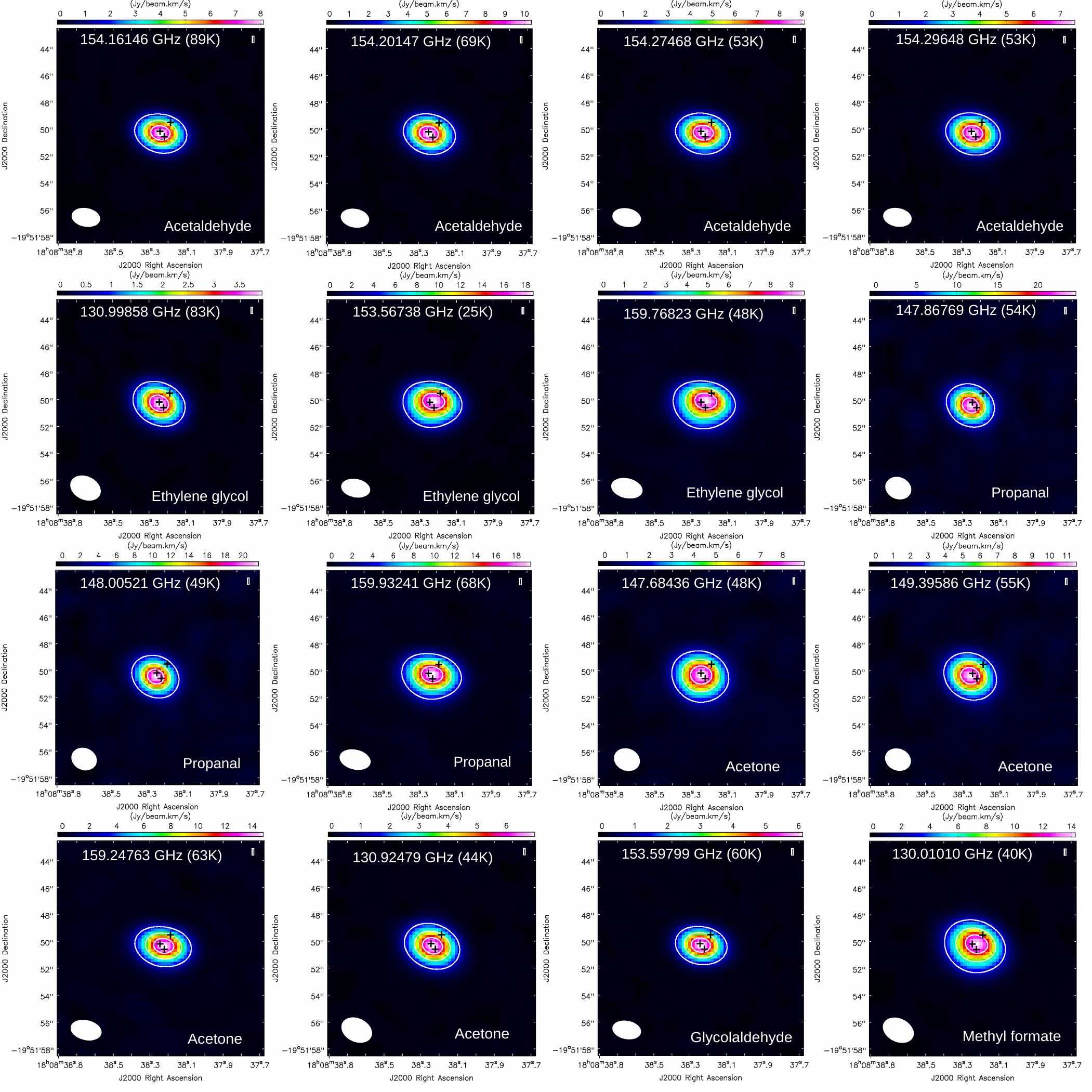}
\caption{Integrated emission maps of observed transitions with upper energy state less than 100 K. Name of the molecules along with their respective transitions (in GHz) and upper energy state (in K) are given in each panel. The contours are shown at $20,\ 50$, and $80\%$ of maximum flux. The observed beam is shown in the lower-left corner of each figure. Three blue plus signs indicate the HII regions B1, B2, and A in the anticlockwise direction starting from the left black cross situated in the white and pink zones of continuum images \citep{cesa10}.}
\label{fig:maps1}
\end{figure*}

\begin{figure*}
\centering
\includegraphics[width=18cm,height=10cm]{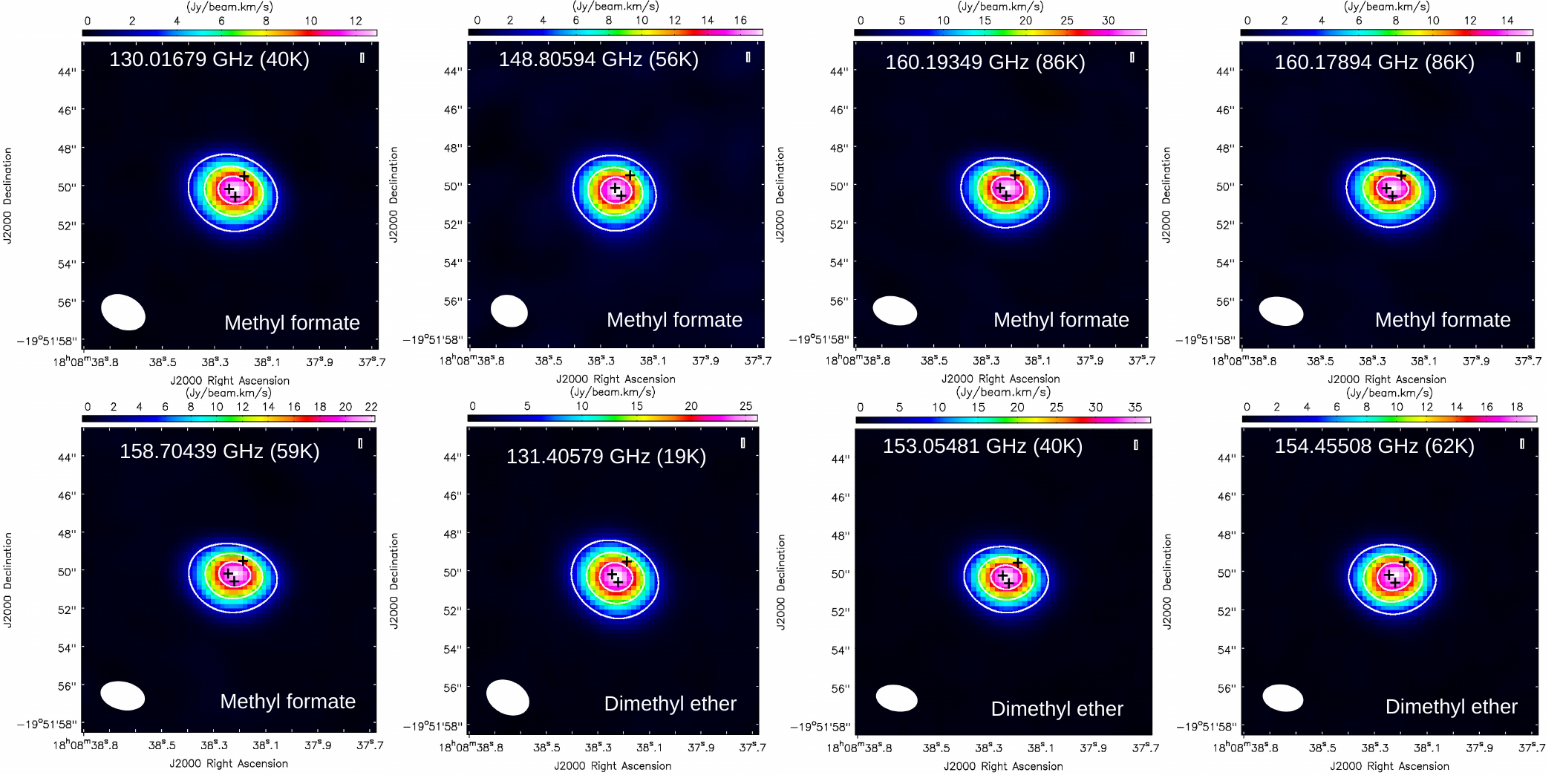}
\caption{ Same as Figure \ref{fig:maps1}.}
\label{fig:maps2}
\end{figure*}

\begin{figure*}
\centering
\includegraphics[width=18cm,height=20cm]{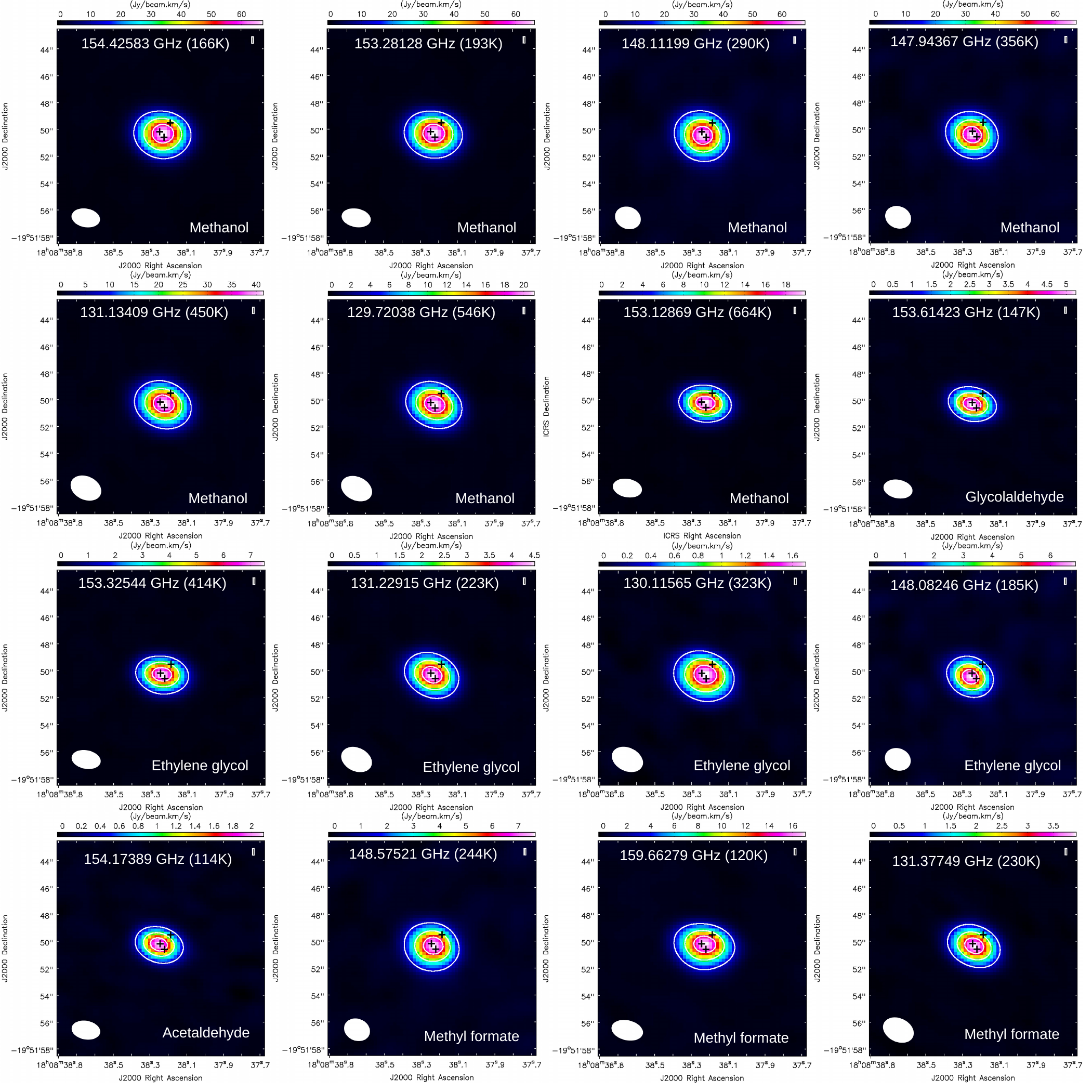}
\caption{Integrated emission maps of observed transitions with upper energy state greater than 100 K are shown.  The contours are shown at $20,\ 50$, and $80\%$ of maximum flux. The observed beam is shown in the lower-left corner of each figure. Three blue plus signs indicate the HII regions B1, B2, and A in the anti-clockwise direction starting from the left black cross situated in the white and pink zones of continuum images \citep{cesa10}.}
\label{fig:maps3}
\end{figure*}

\begin{figure*}
\centering
\includegraphics[width=18cm,height=10cm]{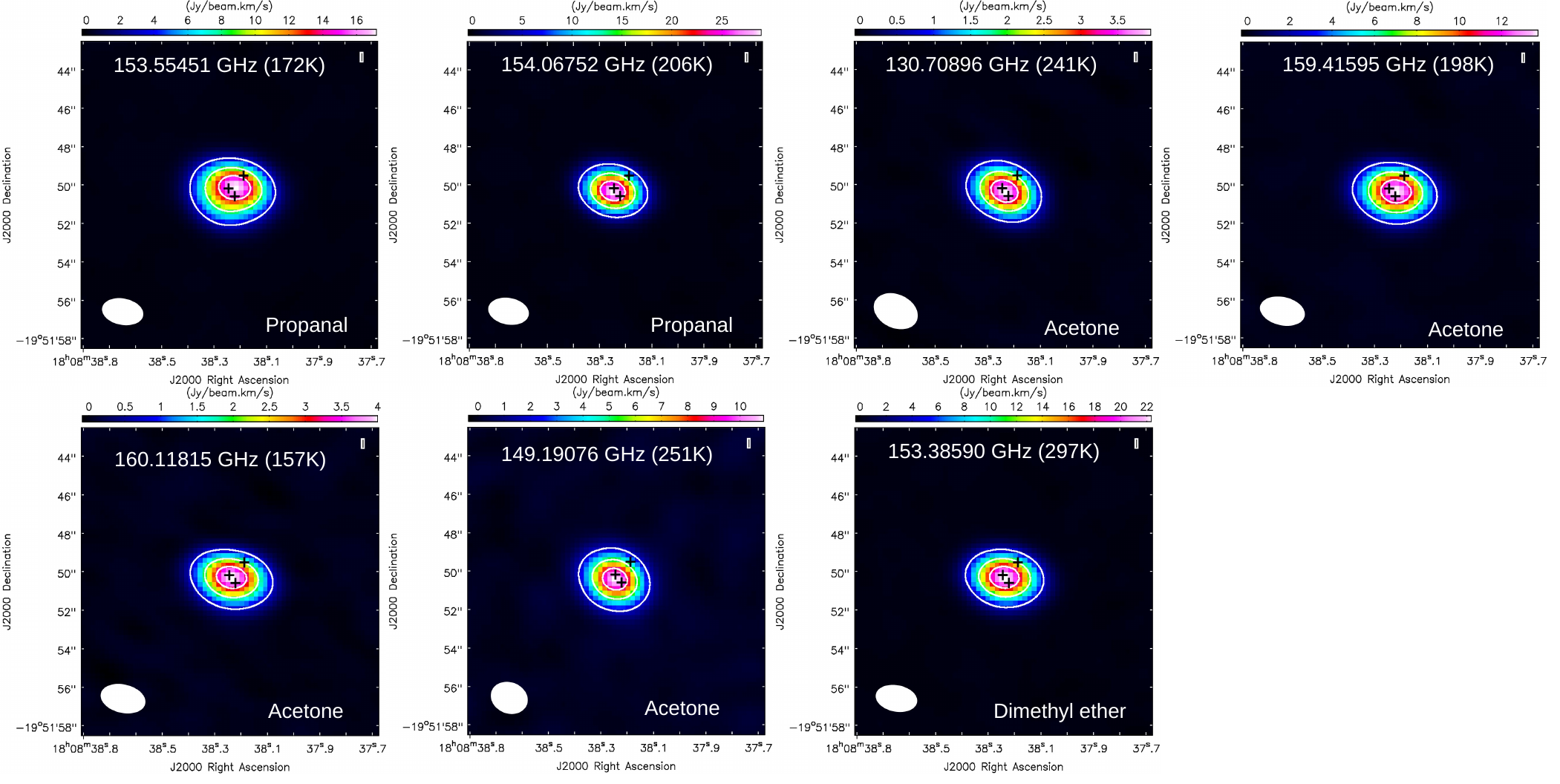}
\caption{Same as Figure \ref{fig:maps3}.}
\label{fig:maps4}
\end{figure*}

 The integrated intensity distribution of all observed molecules is shown in Figures \ref{fig:maps1}, \ref{fig:maps2}, \ref{fig:maps3}, and  \ref{fig:maps4}. The integrated intensity is obtained in the velocity range where the emission line is seen. It is typically obtained between 58 km/s to 78 km/s. This range slightly varies for some species. Figures \ref{fig:maps1} and \ref{fig:maps2} show the intensity distribution of the observed molecules having upper energy state $<100$ K and Figures \ref{fig:maps3} and \ref{fig:maps4}
show the intensity distribution for the molecules with upper energy state $>100$ K.

\clearpage
\bibliography{ald-al-paper}{}
\bibliographystyle{aasjournal}

\end{document}